\documentclass[journal]{IEEEtran}
\usepackage{cite}

\usepackage[pdftex]{graphicx}

\usepackage{amsmath}

\usepackage{algorithmic}

\usepackage{url}

\usepackage{bm}

\usepackage{algorithm}

\usepackage{amssymb}

\def\E{\textbf{E}}
\def\C{\textbf{C}}
\def\D{\textbf{D}}

\usepackage{multirow}
\usepackage{makecell}
\newcommand\toprule{\Xhline{.10em}}
\newcommand\midrule{\Xhline{.08em}}
\newcommand\bottomrule{\Xhline{.10em}}

\usepackage{bbm}

\begin{document}

\title{Non-Parallel Sequence-to-Sequence Voice Conversion with Disentangled Linguistic and Speaker Representations}

\author{Jing-Xuan Zhang,
        Zhen-Hua Ling,~\IEEEmembership{Senior Member,~IEEE},
        and~Li-Rong Dai     
\thanks{
This work was supported by National Key R\&D Program of China (Grant
No. 2019YFF0303001), the National Nature Science Foundation of China
(Grant No. 61871358) and the Key Science and Technology Project of Anhui
Province (Grant No. 18030901016).

J.-X. Zhang, Z.-H. Ling and L.-R. Dai are with the National Engineering Laboratory for Speech and Language Information Processing,
University of Science and Technology of China, Hefei, 230027, China (e-mail: nosisi@mail.ustc.edu.cn, zhling@ustc.edu.cn, lrdai@ustc.edu.cn).
}}

\markboth{PREPRINT MANUSCRIPT OF IEEE/ACM TRANSACTIONS ON AUDIO, SPEECH AND LANGUAGE PROCESSING \copyright2019 IEEE}%
{Jing-Xuan Zhang \MakeLowercase{\textit{et al.}}: Non-parallel Sequence-to-Sequence Voice Conversion with Disentangled Linguistic and Speaker Representations}

\maketitle

\begin{abstract}
This paper presents a method of sequence-to-sequence (seq2seq) voice conversion using non-parallel training data.
In this method, disentangled linguistic and speaker representations
are extracted from acoustic features,
and voice conversion is achieved by preserving the linguistic representations of source utterances while replacing the speaker representations with the target ones.
Our model is built under the framework of encoder-decoder neural networks. 
A recognition encoder is designed to learn the disentangled linguistic representations with two strategies.
First, phoneme transcriptions of training data are introduced to provide the references for leaning linguistic representations of audio signals.
Second, an adversarial training strategy is employed to further wipe out speaker information from the linguistic representations. 
Meanwhile, speaker representations are extracted from audio signals by a speaker encoder. 
The model parameters are estimated by two-stage training, including a pre-training stage using a multi-speaker dataset and a fine-tuning stage using the dataset of a specific conversion pair.
Since both the recognition encoder and the decoder for recovering acoustic  features are seq2seq neural networks,
there are no constrains of frame alignment and frame-by-frame conversion in our proposed method.
Experimental results showed that our method obtained higher similarity and naturalness than the best non-parallel voice conversion method in Voice Conversion Challenge 2018.
Besides, the performance of our proposed method was closed to the state-of-the-art parallel seq2seq voice conversion method. 
\end{abstract}


\section{Introduction}

\IEEEPARstart{V}{oice} conversion (VC) aims at converting the input speeches of a source speaker to make it as if uttered by a target speaker without altering the linguistic content \cite{Childers1985Voice,Childers1989Voice}. Voice conversion has wide applications such as personalized text-to-speech synthesis, entertainment, security attacking and so on \cite{Kain1998Spectral,Arslan1999Speaker,1643640}.

The data conditions for VC can be divided into parallel and non-parallel ones \cite{Mohammadi2017An}.
Parallel VC methods are designed for the datasets with utterances of the same linguistic content but uttered by different persons.
Thus, acoustic models that map the acoustic features of source speakers to those of target speakers can be learned directly when they are aligned.
The forms of the acoustic models for VC included
joint density Gaussian mixture models (JD-GMMs) \cite{Kain1998Spectral,2007Dynamic,Toda2007Voice}, deep neural networks (DNNs) \cite{Desai2009voice,Desai2010Spectral,Chen2014Voice}, recurrent neural networks (RNNs)\cite{Sun2015Voice,nakashika2015voice}, and so on. 
Recently, sequence-to-sequence (seq2seq) neural networks \cite{sutskever2014sequence, cho2014learning, bahdanau2015neural, luong2015effective} have also been applied to VC, which achieved higher naturalness and similarity  than
conventional frame-aligned conversion \cite{8607053, Tanaka2019, zhang2019improving}.

\begin{figure*}[!t]
  \centering
  \includegraphics[width=0.65\textwidth]{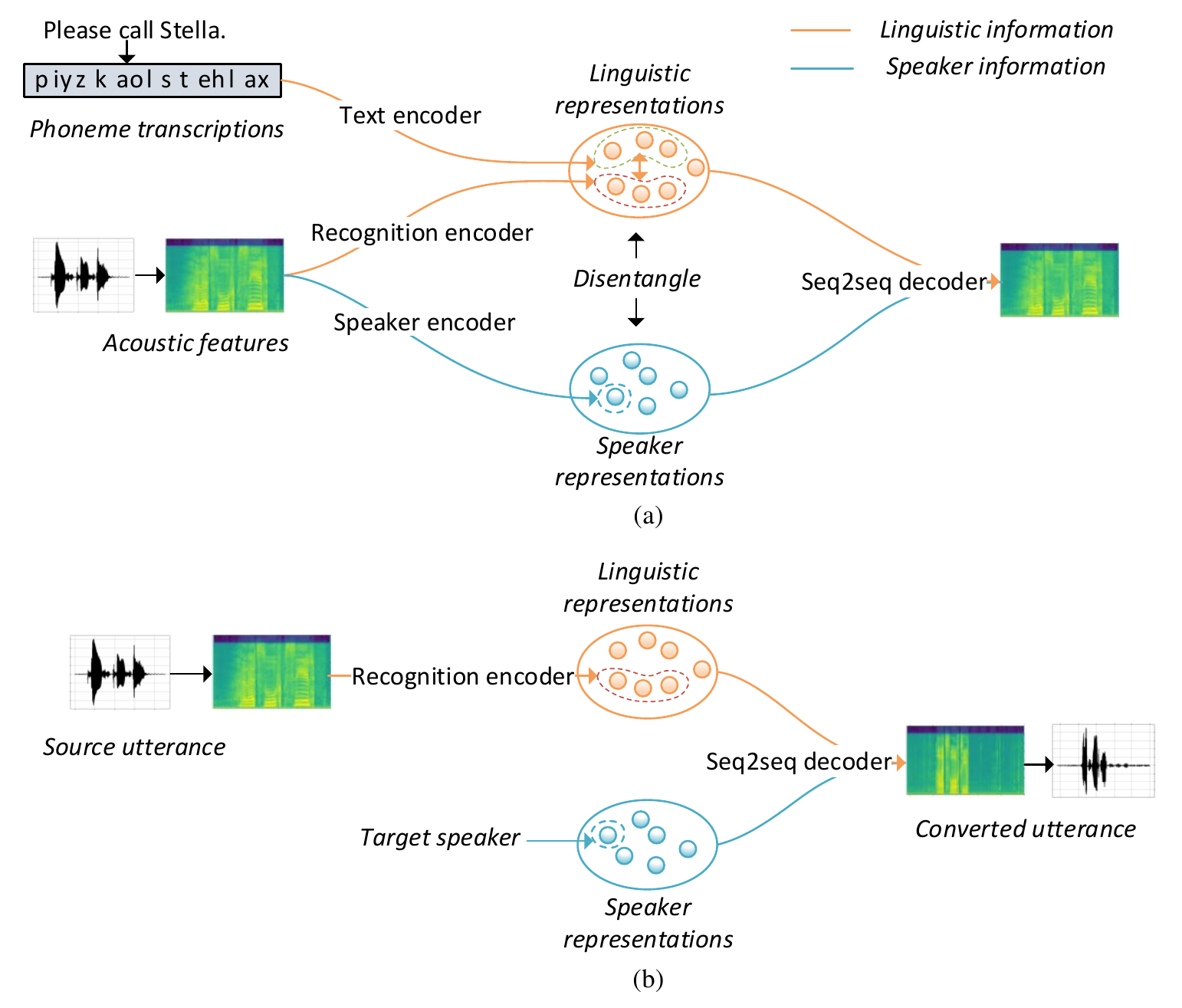}\\
  \centering
  \caption{(a) The overview of our model at the training stage and (b) the conversion process of our proposed method. }\label{fig:fig1}
\end{figure*}

Non-parallel VC is more challenging but more valuable in practice considering the difficulty of collecting parallel training data of different speakers. 
The methods for non-parallel VC can be roughly divided into two categories. The methods of the first category handle non-parallel VC by first converting it into the parallel situation and then learning the mapping functions, such as generating parallel data through text-to-speech synthesis (TTS) \cite{duxans2006voice}, frame-selection \cite{sundermann2006text}, iterative combination of a nearest neighbor search step and a conversion step alignment (INCA) \cite{erro2007frame,erro2010inca} and CycleGAN-based VC \cite{kaneko2017parallel,fang2018high, kaneko2019parallel}. On the other hand,
the methods of the second category factorize the linguistic and speaker related representations
 carried by acoustic features \cite{nakashika2016non,sun2016phonetic,miyoshi2017voice,ljliu2018wav,liu2018voice,saito2018non,  hsu2016voice,hsu2017voice,chou2018multi}.
At the conversion stage, the linguistic content of the source speaker is preserved while the speaker representation of the source speaker is transformed to that of the target speaker.
In contrast, the parallel VC does not need to perform such factorization explicitly.
For a pair of aligned frames, they carry the same linguistic content.
Therefore, the mapping function between them can achieve the transformation of speaker representations.

One representative approach of the second category mentioned above is the recognition-synthesis approach to non-parallel VC \cite{sun2016phonetic,miyoshi2017voice,ljliu2018wav,liu2018voice}.
Typically, it concatenates an automatic speech recognition (ASR) model for extracting linguistic information, such as the posterior probabilities or bottleneck features of phoneme classification,
and a speaker-dependent synthesis model for generating voice of the target speaker.
Despite its success, conventional recognition-synthesis methods have several deficiencies.
First, an extra ASR model is required for
extracting linguistic descriptions. This model is usually trained alone without joint optimization with the synthesis model.
Second, the ASR model is usually trained with a phoneme classification loss and lacks explicit consideration on disentangling linguistic and speaker representations.
Third, most of these methods follow the framework of frame-by-frame conversion and can not achieve the
advantages of seq2seq modeling \cite{8607053}, such as duration modification.

Therefore, a non-parallel seq2seq VC method with disentangled linguistic and speaker representations is presented in this paper.
In this method, a seq2seq recognition encoder and a neural-network-based speaker encoder are constructed for transforming acoustic features into disentangled linguistic and speaker representations.
A seq2seq decoder is built for recovering acoustic features from the combination of them. 
\figurename~\ref{fig:fig1} (a) depicts the overview of our model at the training stage
and \figurename~\ref{fig:fig1} (b) shows the conversion process of our proposed method, where a WaveNet vocoder is adopted \cite{denoord2016wavenet} for waveform reconstruction.

As shown in \figurename~\ref{fig:fig1} (a), two strategies are proposed to learn the speaker-irrelevant linguistic representations.
First, phoneme transcriptions of audio signals are sent into a text encoder and the outputs are adopted
as the references for learning linguistic representations from acoustic features.
Second, an adversarial training strategy is further designed for eliminating speaker-related information from  the linguistic representations.
The model parameters are estimated by two-stage training, including pre-training using a multi-speaker dataset and fine-tuning on the dataset of a specific conversion pair.
As shown in \figurename~\ref{fig:fig1} (b),
the conversion stage includes first extracting linguistic representations from the source utterance and then
reconstructing acoustic features from them together with the speaker representations of the target speaker.
The text inputs are only used at training time and the conversion process does not rely on any text inputs.

Experiments have been conducted to compare our proposed method with state-of-the-art parallel and non-parallel VC methods objectively and subjectively.
The results showed that our proposed method achieved higher similarity and naturalness than the best non-parallel VC method in Voice Conversion Challenge 2018 (VCC2018). 
Besides, its performance was close to the state-of-the-art parallel seq2seq VC method. 
Some ablation tests have also been conducted to confirm the effectiveness of our proposed method.

\section{Related Work}

\subsection{Recognition-synthesis approach to non-parallel VC}


Sun \emph{et al.} \cite{sun2016phonetic} proposed to extract phonetic
posteriorgrams (PPGs) from source speech using an ASR model then feed them into a deep bidirectional long short-term memory (BLSTM) model  \cite{hochreiter1997long}
for generating converted speech.
Miyoshi \emph{et al.} \cite{miyoshi2017voice} proposed a seq2seq learning method for converting context posterior probabilities, which included a recognition model and a synthesis model.
An any-to-any voice conversion framework was proposed based on a multi-speaker synthesis model conditioned on the $i$-vectors and the outputs of an ASR model \cite{liu2018voice}.
In the study of Liu \emph{et al.} \cite{ljliu2018wav}, the ASR model was estimated using a large-scale training set and WaveNet vocoders were built with limited training data of target speakers for waveform recovery.
This method achieved the best performance of non-parallel VC in Voice Conversion Challenge 2018.

Compared with Miyoshi's method \cite{miyoshi2017voice}, the method proposed in this paper does not use a separate conversion model for converting linguistic representations. 
In contrast, we assume a uniform linguistic space across speakers. 
The recognition encoder compresses acoustic features into linguistic representations which have equal lengths with phoneme transcriptions.
Compared with other recognition-synthesis based VC methods \cite{sun2016phonetic, liu2018voice,ljliu2018wav}, the recognition encoder and the seq2seq decoder in our model are optimized jointly. Disentangled linguistic and speaker representations are also proactively learned in our proposed method.

\def\t{\bm{t}}
\def\h{\bm{h}}
\def\a{\bm{a}}
\def\H{\bm{H}}
\def\T{\bm{T}}
\def\A{\bm{A}}

\begin{figure*}
  \centering
  \includegraphics[width=0.7\textwidth]{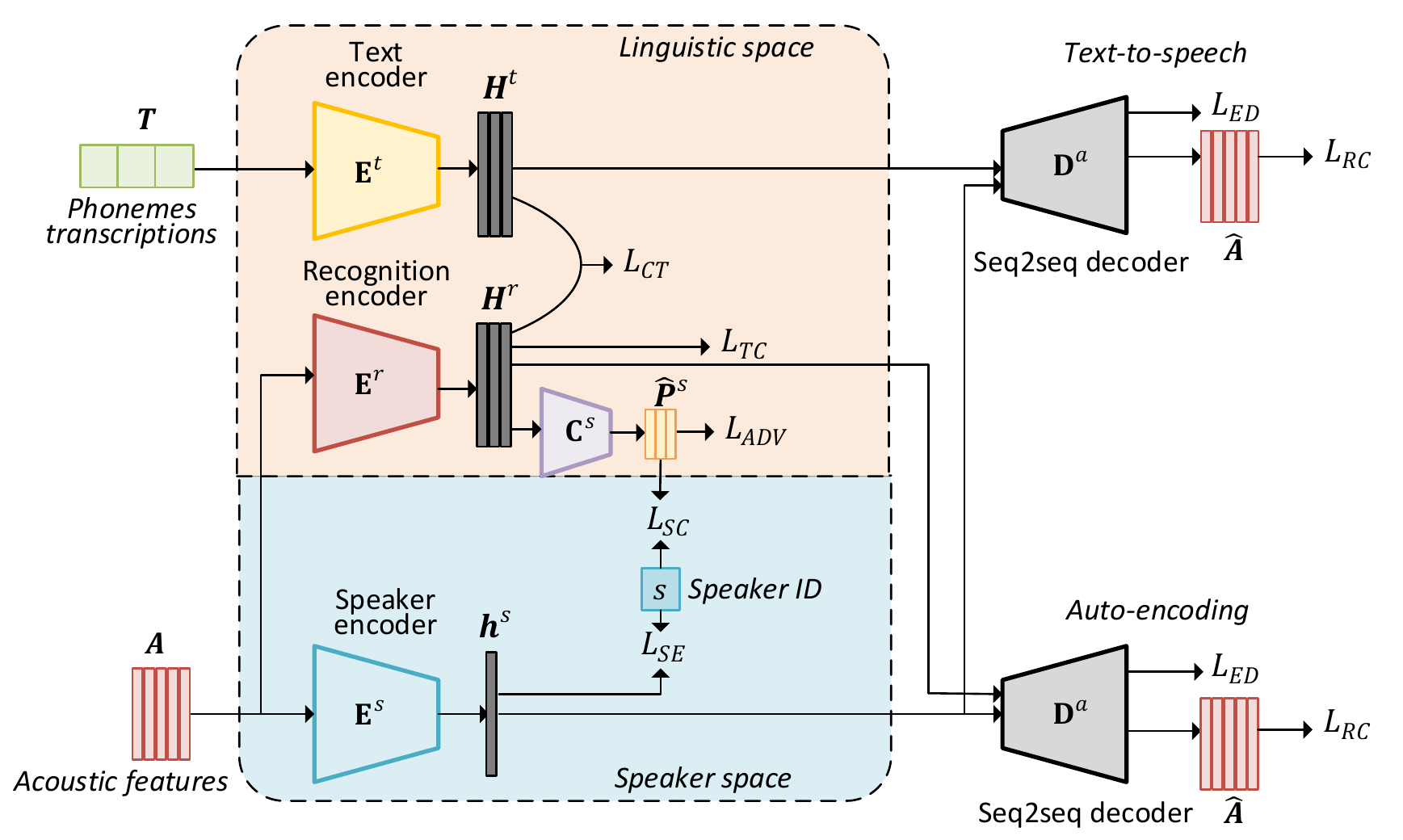}\\
  \centering
  \caption{The architecture of our proposed model and its forward propagation paths during training. 
  The seq2seq decoder adopts the output of either recognition encoder or text encoder as input at each training step. 
   $\H^t$, $\H^r$ and $\h^s$ represent the linguistic embedding from text, the linguistic embedding from audio and the speaker embedding respectively.}\label{fig:fig2}
\end{figure*}


\subsection{Auto-encoder based voice conversion}

The VC methods using auto-encoders (AEs) and variational auto-encoders (VAEs) \cite{hsu2016voice, hsu2017voice} have also been studied in recent years.
Saito \emph{et al.} \cite{saito2018non} proposed to use PPGs for improving VAE-based VC.
Several studies proposed AE-based VC with adversarial learning of hidden representations against speakers information \cite{chou2018multi,polyak2019,ocal2019}.
Polyak \emph{et al.}  \cite{polyak2019} tried to incorporate an attention module between the encoder and the decoder in a WaveNet-based AE.
However, it degraded the mean opinion score (MOS) in evaluation.

Compared with the unsupervised learning of hidden representations in AE or VAE based VC,
our method employs the supervision of corresponding phoneme transcriptions together with adversarial training to learn the recognition encoder.
Furthermore, in contrast to the frame-level encoders and decoders in most previous studies,
the joint training of the recognition encoder and the decoder in our proposed method can be viewed as building a sequence-level auto-encoder.

\subsection{Voice cloning}
Voice cloning is a task that learns the voice of unseen speakers from a few speech samples for text-to-speech synthesis \cite{arik2018neural,jia2018transfer,nachmani2018fitting}.  
Voice cloning takes texts as model inputs, which contain only linguistic information. 
In contrast,  audio signals are used as the inputs of the VC task, which contain not only linguistic content but also speaker identity.
Therefore, carefully disentangling acoustic features into linguistic and speaker representations
is important for achieving high-quality VC in our proposed method. 
It is also possible to incorporate the techniques developed for voice cloning, such as the method of estimating speaker embeddings with limited data,
into our proposed method for achieving the one-shot or few-shot learning of VC.


\section{Proposed Method}
\label{sec:proposed}

\subsection{Model architecture}


 The proposed model contains five components, including a text encoder $\E^{t}$, a recognition encoder $\E^{r}$, a speaker encoder $\E^{s}$, 
 an auxiliary classifier $\C^{s}$, 
 and a seq2seq decoder network  $\D^{a}$. 
 The overall architecture of the model is presented in \figurename~\ref{fig:fig2}
 and functions of these components are described as follows.

\textbf{Text encoder $\E^t$:} Text encoder transforms the text inputs into linguistic embeddings
as ${\H^t} = \E^t(\T)$, where ${\bm{T}}=[\t_1,\dots,\t_N]$ denotes the transcription sequence with one-hot encoding for each phoneme and
$\H^t=[\h^t_1,\dots, \h^t_N]$ denotes the sequence of embedding vectors. $N$ represents the length of the phoneme sequence and the embedding sequence.
The text encoder is built with stacks of convolutional layers followed by a BLSTM and a fully connected layer on the top.

\textbf{Recognition encoder $\E^r$:} Recognition encoder accepts the acoustic feature sequence $\A=[\a_1, \dots, \a_M]$ as inputs
 and predicts the phoneme sequence $\bm{T}$, where $M$ represents the number of acoustic frames.
 The outputs of hidden units before the softmax layer are extracted as
$ \H^r = \E^r(\A)$,
 where $\H^r=[\h^r_1,\dots,\h^r_N]$ denotes the linguistic representations of audio signals.
The recognition encoder $\E^r$ is a seq2seq neural network which aligns the acoustic and phoneme sequences automatically.
Its encoder is based on pyramid BLSTM \cite{chan2016listen} and its attention-based decoder is one-layer LSTM.
Since one phoneme usually corresponds to tens of acoustic frames, we have $M>>N$ and the encoding is a \emph{compression} process.
At the training stage, the output of the recognition encoder $\H^r$  has the equal length to the phoneme sequence $\bm{T}$ regardless of the speaking rate of speakers.
$\H^r$ is expected to reside in the same linguistic space as $\H^t$ and contains only information of linguistic content.

\textbf{Speaker encoder $\E^s$:} The speaker encoder embeds the acoustic feature sequence into a vector as $\h^s = \E^s(\A)$, which can discriminate speaker identities.
The speaker embedding should contain only speaker-related information.
Our speaker encoder is built with stacks of BLSTM followed by an average pooling
and a fully connected layer. The speaker encoder is only employed at the pre-training stage
which will be introduced in Section \ref{model_training}. 
 At the beginning of fine-tuning stage, a trainable speaker embedding is introduced for each speaker and is initialized by
 the $\h^s$ extracted by the speaker encoder.

\textbf{Auxiliary classifier $\C^s$:} The auxiliary classifier is employed to predict the speaker identity from the linguistic representation of the audio input as $\hat{\bm{P}}^s = \C^s({\H^r})$, where $\hat{\bm{P}}^s = [\hat{\bm{p}}^s_1, \dots, \hat{\bm{p}}^s_N]$ and each element $\hat{\bm{p}}^s_n$ is the predicted probability distribution among speakers.
$\C^s$ is introduced for  adversarial training in order to further eliminate speaker-related information remained within the linguistic representation $\H^r$.
In our implementation, $\C^s$ is a DNN which makes prediction for each input embedding vector.


\textbf{Seq2seq decoder $\D^a$:} The seq2seq decoder recovers the acoustic feature sequence from the combination of linguistic embeddings and speaker embeddings as $\hat{\bm{A}} = \D^a({\h^s}, {\H^t})$ or
$\hat{\A} = \D^a({\h^s}, {\H^r})$.
$\hat{\A} = [\hat{\bm{a}}_1,\dots, \hat{\bm{a}}_M]$ represents
 the reconstructed acoustic features
 and either $\H^t$ or $\H^r$ is fed into the decoder at each training step, in which condition
  a process of text-to-speech or auto-encoding of acoustic features is performed.
 It can be viewed as a \emph{decompressing} process in which the linguistic contents are transformed back into acoustic features conditioned on the speaker identity information.
 Here, the structure of the seq2seq decoder is similar to the Tacotron model~\cite{wang2017tacotron:, shen2017natural} for speech synthesis. 



\subsection{Loss functions for disentangled linguistic representations}
Three loss functions are designed for extracting the disentangled linguistic representations from audio signals and their details are as follows.
\subsubsection{Phoneme sequence classification}
The recognition encoder is a seq2seq transducer that
maps input acoustic feature sequences into the sequences of linguistic representations.  
The phoneme classification loss of the linguistic representation sequence $\bm{H}^r$ is defined as
\begin{equation}
\mathcal{L_{TC}}= \frac{1}{N}\sum_{n=1}^{N} \text{CE} (  \t_n, \text{softmax}(\bm{W}{\h^r_n)) )},
\label{ep1}
\end{equation}
where CE$(\cdot)$ represents the cross entropy loss function, $\bm{W}$ is a trainable weight matrix of $\E^r$,
$\h^r_n$ and $\t_n$ denote the linguistic representation and the true label of the $n$-th phoneme respectively.

\subsubsection{Embedding similarity with text inputs}
The linguistic representations extracted from audio signals and from phoneme sequences (i.e., ${\bm{H}^r}$ and ${\bm{H}^t}$) are expected to 
share the same linguistic space. 
Intuitively, we would like the linguistic embeddings from both audio and text inputs
 to be similar with close distance.
Inspired by previous studies on feature mapping \cite{chopra2005learning}, lip sync \cite{chung2016out} and learning joint embedding space from audio and video inputs \cite{zhou2018talking},
  the contrastive loss is adopted in this paper to increase the similarity between $\bm{h}^r_m$ and $\bm{h}^t_n$ if $m=n$ and to reduce their similarity if $m\neq n$.

 The loss function is defined as
\begin{equation}
\mathcal{{L}_{CT}} = \sum_{m=1,n=1}^{N,N}{\mathbb{I}_{mn}d_{mn} +
(1-\mathbb{I}_{mn})\text{max}(1-d_{mn}, 0)}.
\label{eq2}
\end{equation}
$\mathbb{I}_{mn}$ is the element of an indicator matrix where $\mathbb{I}_{mn}=1 $ if $m=n$ and $\mathbb{I}_{mn}=0$ otherwise.
$d_{mn}$ is the distance between $\bm{h}^r_m$ and $\bm{h}^t_n$ which is defined as
\begin{equation}
d_{mn}=\parallel \frac{\bm{h}^r_m}{\parallel \bm{h}^r_m \parallel _2} -
\frac{\bm{h}^t_n}{\parallel \bm{h}^t_n \parallel _2} \parallel ^2_2.
\label{eq4}
\end{equation}
In our experiments, we found that 
the second term of the left part in Eq.~(\ref{eq2}) was necessary, which prevented 
the extracted representations from falling into the same vector. 

\subsubsection{Adversarial training against speaker classification}
The auxiliary classifier $\C^s$ is trained with a cross entropy loss  $\mathcal{L_{SC}}$
between the predicted speaker probabilities and the target labels as
\begin{equation}
\mathcal{L_{SC}}= \frac{1}{N}\sum_{n=1}^{N}{ \text{CE} (\bm{p}^s , \hat{\bm{p}}^s_n )},
\end{equation}
where $\bm{p}^s$ is the one-hot speaker label of input audio signals. 

Meanwhile, the recognition encoder $\E^r$ is optimized toward the opposite goal, i.e., fooling the auxiliary classifier to
make a prediction of equal probabilities  among speakers.
Thus, an adversarial loss $\mathcal{L_{ADV}}$ is designed for training $\E^r$ as
\begin{equation}
\mathcal{L_{ADV}} = \frac{1}{N}\sum_{n=1}^{N}{ \parallel 
\bm{e} - \hat{\bm{p}}^s_n \parallel ^2_2}.
\label{eq5}
\end{equation}
where $\bm{e} = [1/S, \dots, 1/S]^\top$ is an uniform distribution and $S$ is the total number of speakers.
When updating the parameters of the recognition encoder, $\C^s$ is frozen. 
It is expected to reduce the speaker-related information carried by the linguistic representations of audio signals by minimizing $\mathcal{L_{ADV}}$.
If the speaker representations are completely
eliminated from linguistic hidden embeddings, the auxiliary classifier should
achieve the minimum loss and
assign equal probability to each possible speaker.
Similar loss functions have been used for disentangling person identity and word space of videos \cite{zhou2018talking}.

\subsection{Loss functions for disentangled speaker representations}

In parallel to producing linguistic representations, speaker embeddings are extracted from acoustic features by the speaker encoder $\E^s$.
Speaker embeddings are expected to be discriminative to the speaker identity. Therefore, we introduce
a speaker classification loss for the training of $\E^s$.
The speaker classification loss of $\E^s$ is calculated as
\begin{equation}
\mathcal{L_{SE}} = \text{CE} (  \bm{p}^s , \text{softmax}(\bm{V}\bm{h}^s)).
\end{equation}
where $\bm{V}$ is a trainable weight matrix of $\E^s$.  
Once the speaker representation $\bm{h}^s$ for an input utterances is obtained, it is processed by $L_2$ normalization and fixed 
when passed through the decoder.
In another word, the speaker encoder is only optimized by $\mathcal{L_{SE}}$ 
and not influenced by further calculations using $\bm{h}^s$.
Based on our experiments, this loss function can help to obtain consistent speaker embeddings from different utterances of the same speaker.
Hence, we do not conduct adversarial training for extracting speaker embeddings. 

\subsection{Loss functions for acoustic feature prediction}

Acoustic features are eventually recovered from the linguistic representations $\bm{H}^r$ or $\bm{H}^t$
together with the speaker embedding $\bm{h}^s$ via the seq2seq decoder. 
After $L_2$ normalization, the $\bm{h}^s$ vector is concatenated with the linguistic representation of each phoneme. 
An $L_1$ loss is defined for the predicted acoustic features as
\begin{equation}
\mathcal{L_{RC}}= \frac{1}{M}\sum_{i=1}^{M}{
\parallel \hat{\bm{a}}_i - \bm{a}_i \parallel _1},
\label{eq7}
\end{equation}
where $\hat{\bm{a}}_i$ is the predicted acoustic feature vector at the $i$-th frame.

In order to end the acoustic feature sequences generated by the seq2seq decoder at the conversion stage, the hidden state of the seq2seq decoder at each frame is projected to a scalar followed by sigmoid activation
to predict whether current frame is the last frame in an utterance. Accordingly, a cross entropy loss $\mathcal{L_{ED}}$ is defined for this prediction at the training stage.

\begin{algorithm}[!t]
\caption{Pre-training  using a  dataset of $S$ speakers.}
\label{alg1}
\begin{algorithmic}

\REQUIRE~~\\

$\bm{\theta}_{E^t}, \bm{\theta}_{E^r}, \bm{\theta}_{E^s}, \bm{\theta}_{C^s}, \bm{\theta}_{D^a}$, $iter \leftarrow 1.$

\ENSURE~~\\
\WHILE {not converaged}
\STATE Sample mini batch $ \langle \bm{A},\bm{T},\bm{p}^s \rangle $ 
\STATE  $\bm{H}^t \gets \E^t(\bm{T})$, $\bm{H}^r \gets \E^r(\bm{A})$, $\bm{h}^s \gets \E^s(\bm{A})$
\STATE $\hat{\bm{P}}^s \gets \C^s(\H^r)$
     \IF{$iter$ is even}
     \STATE   $\bm{\hat{A}} \leftarrow \D^a(\bm{h}^s, \bm{H}^t)$
     \ELSE
     \STATE   $\bm{\hat{A}} \leftarrow \D^a(\bm{h}^s, \bm{H}^r)$
     \ENDIF
\STATE computing $\mathcal{L_{TC}}, \mathcal{L_{CT}}, \mathcal{L_{ADV}}, \mathcal{L_{SC}},  \mathcal{L_{SE}}, \mathcal{L_{RC}}, \mathcal{L_{ED}}$

     \IF{$iter$ is even}
     \STATE $\bm{\theta}_{E^t} \xleftarrow{+} - \nabla_{\bm{\theta}_{E^t} }(w_{ct}\mathcal{L_{CT}} + \mathcal{L_{RC}} + \mathcal{L_{ED}} )$
     \STATE $\bm{\theta}_{E^r} \xleftarrow{+} - \nabla_{\bm{\theta}_{E^r} }{(  \mathcal{L_{TC}} + w_{ct}\mathcal{L_{CT}} + w_{adv}
\mathcal{L_{ADV}})}$
     \ELSE
     \STATE $\bm{\theta}_{E^t} \xleftarrow{+} - \nabla_{\bm{\theta}_{E^t} }{w_{ct}\mathcal{L_{CT}}}$
     \STATE $\bm{\theta}_{E^r} \xleftarrow{+} - \nabla_{\bm{\theta}_{E^r} }{( \mathcal{L_{TC}} + w_{ct}\mathcal{L_{CT}} + w_{adv}\mathcal{L_{ADV}} + }$
     \STATE   \quad \quad \quad \quad  \ \   $\mathcal{L_{RC}} + \mathcal{L_{ED}})$
     \ENDIF
\STATE $\bm{\theta}_{E^s} \xleftarrow{+} - \nabla_{\bm{\theta}_{E^s}}{\mathcal{L_{SE}}}$
\STATE $\bm{\theta}_{C^s} \xleftarrow{+} - \nabla_{\bm{\theta}_{C^s}}{w_{sc}\mathcal{L_{SC}}}$
\STATE $\bm{\theta}_{D^a} \xleftarrow{+} - \nabla_{\bm{\theta}_{D^a}}{(\mathcal{L_{RC}} +
\mathcal{L_{ED}})}$
\STATE $iter \xleftarrow{+} 1$
\ENDWHILE
\end{algorithmic}
\end{algorithm}

\subsection{Model training}
\label{model_training}
 In summary, there are totally 7 losses introduced above for training our proposed model.
 They are the loss for phoneme sequence classification $\mathcal{L_{TC}}$,
 the contrastive loss for embedding similarity with text inputs $\mathcal{L_{CT}}$,
 the losses for adversarial training $\mathcal{L_{ADV}}$ and $\mathcal{L_{SC}}$,
 the loss for speaker representations $\mathcal{L_{SE}}$, and the losses for
 predicting acoustic features and utterance ends $\mathcal{L_{RC}}$  and $\mathcal{L_{ED}}$.
 These losses are leveraged through weighting factors to form the complete loss function.
 Weighting factors
 $w_{adv}, w_{ct}, w_{sc}$  are introduced for $\mathcal{L_{ADV}},\mathcal{L_{CT}}, \mathcal{L_{SC}}$ respectively.
For other losses, the weighting factors are set as 1 heuristically.

 The model parameters are estimated by two-stage training. 
At the first stage (i.e., the pre-training stage), the whole model is trained using a large multi-speaker dataset which
contains triplets of text transcriptions, speech waveforms and a speaker identity label for each utterance.
Then, the model parameters are updated on a specific conversion pair of source and target speakers at the second stage (i.e., the fine-tuning stage).
It  should be noticed that our model is capable of performing many-to-many VC
if we simply increase the number of speakers during fine-tuning.
However, we concentrate on the voice conversion between a pair of speakers in this paper.

The algorithm for pre-training is shown in Algorithm~\ref{alg1},
where $\bm{\theta}_{E^t}$,  $\bm{\theta}_{E^r}$,  $\bm{\theta}_{E^s}$, $\bm{\theta}_{C^s}$ and $\bm{\theta}_{D^a}$
denote the parameters of the five model components respectively.
The algorithm for fine-tuning is almost the same as 
Algorithm~\ref{alg1}.
The multi-speaker dataset is replaced by the one containing the source speaker and the target speaker,
and the number of speakers is reset as $S=2$.
Two trainable
speaker embeddings  are introduced for these two speakers.
These two speaker embeddings are initialized as the speaker encoder output  $\h^s$ averaged across training utterances of these two speakers respectively.
Then,  the speaker encoder $\E^s$ is discarded during fine-tuning.
Besides, the softmax layer for multi-speaker classification in the auxiliary classifier is replaced by a sigmoid output layer
for the binary speaker classification.


\section{Experiments}
\subsection{Experiment conditions}
\label{sec:expcond}
One female speaker (slt) and one male speaker (rms) in the CMU ACRTIC dataset \cite{arctic} were used
as the pair of speakers for conversion in our experiments.
For each speaker, the evaluation and test set both contained 66 utterances.
The non-parallel training set for each speaker contained 500 utterances. 
For comparison with parallel VC, 500 parallel utterances were also selected for each speaker to form the parallel training set.
The multi-speaker VCTK dataset \cite{veaux2017cstr} was utilized for model pre-training in our proposed method.
Altogether 99 speakers were selected from the VCTK dataset. For each speaker, 10 and 20 utterances were used for validation and testing respectively,
 and the remaining utterances were used as training samples. The total duration of training samples was about 30 hours.

\begin{table}
\renewcommand\arraystretch{1.5}
\caption{Details of the model configurations.}
\label{tab:tab1}
\begin{tabular}{ccl}
\toprule

 \multicolumn{2}{l}{\multirow{3}{*}{$\E^t$}}   & Conv1D-5-512-BN-ReLU-Dropout(0.5) $\times3$   $\to$ \\
                                                   & & 1 layer BLSTM, 256 cells each direction $\to$ \\
                                                   & & FC-512-Tanh $\to$ $\H^t$ \\
 \cline{1-3}

 \multirow{5}{10pt}{$\E^r$}   & \multirow{3}{35pt}{\textbf{Encoder}}  &2 layer Pyramid BLSTM \cite{chan2016listen}, 256 cells \\
                                &                                        &each direction, i.e. reducing the \\
                                &                                        &sequence time resolution by factor 2. \\
\cline{2-3}
                                &\multirow{3}{35pt}{\textbf{Decoder}}     &1 layer LSTM, 512 cells with \\
                                &                                         &location-aware attention~\cite{chorowski2015attention} $\to$ \\
                                &                                          & FC-512-Tanh $\to$ $\H^r$ \\

\cline{1-3}

\multicolumn{2}{l}{\multirow{2}{*}{$\E^s$}}   & 2 layer BLSTM, 128 cells each direction $\to$ \\
                              &                        &average pooling $\to$ FC-128-Tanh $\to \h^s $ \\
\cline{1-3}

\multicolumn{2}{l}{\multirow{2}{*}{$\C^s$}} & FC-512-BN-LeakyReLU \cite{radford2015unsupervised} $\times3$ $\to$ \\
                                              &     & FC-99-Softmax $\to$ $\bm{\hat{P}}^s$ \\
\cline{1-3}

\multirow{8}{*}{$\D^a$}  & \multirow{1}{35pt}{\textbf{Encoder}} & 1 layer BLSTM, 256 cells each direction \\
\cline{2-3}
                              & \multirow{1}{35pt}{\textbf{PreNet}}                    & FC-256-ReLU-Dropout(0.5) $\times2$ \\
\cline{2-3}
                              & \multirow{3}{35pt}{\textbf{Decoder}}       & 2 layer LSTM, 512 cells with  \\
                              &                                             &forward attention \cite{zhang2018forward},\\
                              &                                             & 2 frames are predicted each decoder step \\
\cline{2-3}
                              & \multirow{3}{35pt}{\textbf{PostNet}}         & Conv1D-5-512-BN-ReLU-Dropout(0.5) $\times5$ $\to$ \\
                              &                          & Conv1D-5-80, with residual connection \\
                              &                          & from the input to output \\

\bottomrule
\multicolumn{3}{p{230pt}}{
``FC'' represents fully connected layer. ``BN'' represents batch normalization.
``Conv1D-$k$-$n$'' represents 1-D convolution with kernel size $k$ and channel $n$. ``$\times N$'' represents
repeating the block for $N$ times. $\D^a$ follows the framework of the Tacotron model \cite{shen2017natural}.}
\end{tabular}
\end{table}

The acoustic features were 80-dimensional Mel-spectrograms extracted every 12.5 ms and the frame size for short-time Fourier transform (STFT) was 50 ms.
The original Mel-spectrograms were then scaled to logarithmic domain.
In order to obtain the inputs of the text encoder, we generated phoneme transcriptions using the grapheme-to-phoneme module of Festival\footnote{\texttt{\url{http://www.cstr.ed.ac.uk/projects/festival/}}.}.
Our model was implemented with PyTorch\footnote{\texttt{\url{https://pytorch.org/}}.}. The Adam optimizer \cite{kingma2014} was used and the training batch size was 32 and 8 at the pre-training and fine-tuning phases respectively. The learning rate was fixed to 0.001 for the 80 epoches of pre-training and it was halved every 7 epoches during fine-tuning.
The weighting factors of loss functions were tuned on the validation set of the multi-speaker data and were determined as $w_{ct}=30$ and $w_{sc}=0.1$. $w_{adv}$ was set as 20 and 0.2 during pre-training and fine-tuning respectively.

The details of our model structures are summarized in \tablename~\ref{tab:tab1}\footnote{Implementation code is available at \texttt{ \url{https://github.com/jxzhanggg/nonparaSeq2seqVC_code/}}.}.
A beam search with width of 10 was adopted for inference using the recognition encoder $\E^r$.
The WaveNet vocoder predicted 10-bit waveforms with $\mu$-law companding.
Its implementation followed our previous work \cite{ljliu2018wav}.

\begin{table*}[t]
 \renewcommand\arraystretch{1.5}
 \caption{Objective evaluation results of different methods.}\label{tab:obj}
 \centering
  \begin{tabular}{c c l c c c c c}
 \toprule
Conversion Pairs &  \multicolumn{2}{c}{Methods}  & MCD (dB)   & $F_0$ RMSE (Hz) & VUV (\%)  &$F_0$ CORR  &DDUR (s)\\
\midrule
 \multirow{5}{*}{rms-to-slt} & \multirow{2}{*}{\emph{para}} & DNN       &4.134  &16.651    &9.205  &0.585    &0.481 \\
 \cline{3-8}
                                &   & Seq2seqVC &\textbf{2.999}  &\textbf{13.633}    &\textbf{6.968}  &\textbf{0.727}    &\textbf{0.152} \\
 \cline{2-8}
                                & \multirow{3}{*}{\emph{non-para}} & CycleGAN  &3.309  &27.264    &11.603  & 0.394  &0.481 \\
 \cline{3-8}
                                &  & VCC2018   &3.376  &\textbf{15.042}    &8.222   & \textbf{0.663}  &0.481 \\
 \cline{3-8}
                                & & Proposed  &\textbf{3.088} & 16.043    &\textbf{7.898}   &0.624  &\textbf{0.261} \\
 \midrule
  \multirow{5}{*}{slt-to-rms}  & \multirow{2}{*}{\emph{para}} & DNN       &3.747  &16.484    &11.750  &0.526    &0.481 \\
 \cline{3-8}
                                & & Seq2seqVC  &\textbf{2.887}  &\textbf{14.360}    &\textbf{9.435}  &\textbf{0.664}    &\textbf{0.245} \\
 \cline{2-8}
                                & \multirow{3}{*}{\emph{non-para}} & CycleGAN  &3.246   &18.284    &13.428  & 0.507  &0.481 \\
 \cline{3-8}
                                & & VCC2018   &3.171  &\textbf{15.771}    &11.382   &\textbf{0.593}  &0.481 \\
 \cline{3-8}
                                &  & Proposed  &\textbf{2.974} & 16.080     &\textbf{10.327}   &0.581  &\textbf{0.264} \\
\bottomrule
  \multicolumn{8}{p{380pt}}{Best results  obtained among parallel and non-parallel VC methods for each metric are highlighted with bold fonts. ``para'' and ``non-para'' represent parallel VC and non-parallel VC respectively.}
  \end{tabular}
\end{table*}

\begin{table*}
\renewcommand\arraystretch{1.5}
\centering
\caption{Mean opinion scores (MOS) with $95\%$ confidence intervals on naturalness and similarity of different methods.}
\label{tab:sub}
\begin{tabular}{c l c c c c}
\toprule
\multicolumn{2}{c}{\multirow{2}{*}{Methods}} & \multicolumn{2}{c}{rms-to-slt} &  \multicolumn{2}{c}{slt-to-rms} \\
\cline{3-6}
                &         & Naturalness       &Similarity       & Naturalness       &Similarlity  \\
\cline{1-6}
\multirow{2}{*}{\emph{para}}&  DNN                      &$2.09 \pm 0.09$ &$2.03 \pm 0.10$    &$2.38 \pm 0.10$  &$2.42 \pm 0.10$ \\
\cline{2-6}
& Seq2seqVC                &$\textbf{4.20} \pm 0.09$  & $\textbf{4.26} \pm 0.09$  &$\textbf{4.18} \pm 0.09$  &$\textbf{4.37} \pm 0.09$ \\
\cline{1-6}
\multirow{3}{*}{\emph{non-para}}& CycleGAN                 &$1.48 \pm 0.09$  &$1.49 \pm 0.08$   &$1.81 \pm 0.11$  &$1.82 \pm 0.11$ \\
\cline{2-6}
&VCC2018                  &$3.53 \pm 0.11$  &$3.59 \pm 0.14$  &$3.76 \pm 0.11$  &$3.89 \pm 0.12$ \\
\cline{2-6}
& Proposed                 &$\textbf{4.19} \pm 0.09$ &$\textbf{4.24} \pm 0.09$  &$\textbf{4.18} \pm 0.09$   & $\textbf{4.26} \pm 0.09$ \\
\bottomrule
\multicolumn{6}{p{300pt}}{Highest scores among parallel and non-parallel VC methods for each metric are highlighted.
}
\end{tabular}
\end{table*}

\subsection{Comparative methods}
Four VC methods were implemented for comparison with our proposed method\footnote{
Audio samples of our experiments are available at \texttt{\url{https://jxzhanggg.github.io/nonparaSeq2seqVC/}}.}. Two of them adopted parallel training and the rest
adopted non-parallel training. The details of them are described as follows.

\textbf{DNN:} Parallel VC method based on a DNN acoustic model.
41-dimensional Mel-cepstral coefficients (MCCs), 5-dimensional  band aperiodicities (BAPs),
1-dimensional fundamental frequency ($F_0$), their delta and accelerate features were extracted as acoustic features.
The Merlin open source toolkit\footnote{\texttt{\url{https://github.com/CSTR-Edinburgh/merlin/}}.} \cite{Zhizheng2016Merlin} was employed for implementation.
The DNN contained 6 layers with 1024 units and \emph{tanh} activations per layer.
WORLD vocoder  \cite{Morise2016WORLD} was adopted for waveform recovery.

\textbf{Seq2seqVC:} Parallel VC method based on a seq2seq model \cite{8607053}. 80-dimensional Mel-spectrogram features were adopted as acoustic features
 together with bottleneck features, which  were linguistic-related descriptions extracted by an ASR model  trained on about 3000 hours of external speech data \cite{8607053}. 
The WaveNet vocoder built in our proposed method was also used here for waveform recovery. 
Previous study showed that this method achieved better performance than the best parallel VC method in VCC2018 \cite{8607053}. 

\textbf{CycleGAN:} Non-parallel VC method based on CycleGAN \cite{kaneko2017parallel}.
An open source implementation of CycleGAN-based VC was adopted\footnote{\texttt{\url{https://github.com/leimao/Voice_Converter_CycleGAN/}}.}.
MCCs, BAPs and $F_0$ were used as acoustic features.
Only MCCs were converted by CycleGAN and $F_0$ trajectories were converted by Gaussion mean normalization \cite{chappell1998speaker-specific}.
The BAP features were not converted.
WORLD vocoder was used for waveform recovery.
Actually, we have tried to adopt the WaveNet vocoder built in our proposed method.
However, the reconstructed voice was noisy and the quality was not as good as that using WORLD vocoder.

\textbf{VCC2018:} Non-parallel VC method based on conventional recognition-synthesis approach \cite{ljliu2018wav}. The ASR model was the same as
the one used by the Seq2seqVC method. Then, bottleneck features were extracted from the built recognition model as linguistic descriptions
and were used as the inputs of speaker-dependent synthesis models.
MCCs, BAPs and $F_0$ features were used as acoustic features and the WaveNet vocoder was adopted for waveform recovery.
This method achieved the best performance on the non-parallel VC task of Voice Conversion Challenge 2018.

\subsection{Objective evaluations}\label{subsubsec:obj}
Mel-cepstrum distortion (MCD), root of mean square errors of
$F_0$ ($F_0$ RMSE), the error rate of voicing/unvoicing flags (VUV) and the Pearson correlation factor of $F_0$ ($F_0$ CORR)
were used as the metrics for objective evaluation.
In order to investigate the effects of
duration modification, we also computed the average absolute differences between the durations of the converted and target utterances (DDUR) as in our previous work \cite{8607053}. When computing DDUR, the silence segments at the beginning and the end of utterances were removed.

Because Mel-spectrograms were adopted as acoustic features in the Seq2seqVC method and our proposed method, it's not straightforward to extract $F_0$ and MCCs features
from the converted acoustic features. Therefore, the MCCs and $F_0$ were extracted from the waveform of converted utterances using STRAIGHT \cite{Kawahara1999Restructuring}. Then, they were aligned to the reference utterances by dynamic time wraping using  MCCs features for calculating the metrics. 

The test set results of both rms-to-slt and slt-to-rms conversions are reported in \tablename~\ref{tab:obj}.
As we can see from this table,
among the parallel VC methods, Seq2seqVC achieved better performance than the DNN method.
For non-parallel VC, our proposed method achieved the best result on MCD, UVU and DDUR metrics.
In terms of $F_0$ RMSE and $F_0$ CORR metrics, the VCC2018 method performed better than our proposed method.
Although there were no parallel training utterances, our proposed method can still reduce the DDUR of the parallel and non-parallel methods following frame-by-frame conversion.
The objective performance of propose method was close to but still not as good as the parallel Seq2seqVC method in spectral and $F_0$  estimation.
For durational conversion, the Seq2seqVC method outperformed other methods by large margins.
The reason is that the Seq2seqVC method made use of the supervision from paired utterances for learning the mapping function at utterance level. While our method can only obtain speaking rate information from the speaker embeddings.
To improve the capability of speaker embeddings on describing
speaking rates is worth further investigation in the future.

\subsection{Subjective evaluations}
Subjective evaluations in term of both the naturalness and similarity of converted speech were conducted.
20 utterances in the test set of each speaker were randomly selected and converted using the five methods mentioned above.
For each utterance, the converted samples were presented to listeners in random order, who were asked to give a 5-scale opinion score (5: excellent, 4: good, 3: fair, 2: poor, 1: bad) on both naturalness and similarity of each sample.
At least thirteen listeners participated in each evaluation and they were asked to use headphones.
The evaluation results are presented in \tablename~\ref{tab:sub}.
As we can see from this table, the Seq2seqVC method and our proposed method achieved the best subjective performance among all parallel and non-parallel methods respectively in both conversion directions.
The DNN and CycleGAN methods obtained lower MOSs than other methods, which was consistent with the results of objective evaluations. 

Although the VCC2018 method adopted a much larger dataset than VCTK for training the recognition model, our method still achieved better performance than it.
In rms-to-slt conversion,
the $p$-values of $t$-tests between these two methods for naturalness and similarity were $7.3\times10^{-22}$ and $3.3\times10^{-19}$ respectively.
In slt-to-rms conversion, the $p$-values for naturalness and similarity were $1.3\times10^{-10}$ and $4.2\times10^{-9}$ respectively.
We can see that the superiority of our proposed method over the VCC2018 method was significant. 

Compared with the parallel Seq2seqVC method, our proposed method achieved close and slightly inferior performance.
In rms-to-slt conversion, the $p$-values for naturalness and similarity were $0.94$ and $0.73$ respectively.
In slt-to-rms conversion, the $p$-values for naturalness and similarity were $0.94$ and $0.03$ respectively.
Therefore, the superior of Seq2seqVC over proposed method is insignificant except the similarity in slt-to-rms conversion.
In addition to using parallel training data, the Seq2seqVC method also benefitted from the  bottleneck features extracted from an ASR model.
Considering that the dataset for training the ASR model was much larger than the VCTK dataset used in our proposed method,
it's possible to further improve our model by adopting a larger multi-speaker dataset for pre-training.

\begin{table*}[t]
 \renewcommand\arraystretch{1.5}
 \caption{Objective evaluation results of our proposed method using different numbers of non-parallel utterances for fine-tuning.}\label{tab:tab4}
 \centering
  \begin{tabular}{c c c c c c c}
 \toprule
 Conversion Pairs              &\# of Utt.& MCD (dB)   & $F_0$ RMSE (Hz) & VUV (\%)  &$F_0$ CORR  &DDUR (s) \\
\midrule
 \multirow{5}{*}{rms-to-slt} &500      &\textbf{3.088}   &16.043   &\textbf{7.898}   &0.624   &\textbf{0.261}\\
 \cline{2-7}
                                 &400      &3.095   &15.544   &8.423    &0.649  &0.263\\
 \cline{2-7}
                                 &300      &3.114   &15.950   &8.037   &0.636  &0.270 \\
 \cline{2-7}
                                 &200      &3.126   &\textbf{15.194}   &7.923   &\textbf{0.670}   &0.286\\
 \cline{2-7}
                                 &100      &3.171   &16.368   &8.410   &0.622   &0.290\\
 \midrule
 \multirow{5}{*}{slt-to-rms} &500      &\textbf{2.974}   &\textbf{16.080}   &\textbf{10.327}  &\textbf{0.581}   &0.264 \\
\cline{2-7}
                                 &400      &3.007   &16.591  & 10.391   & 0.563  &\textbf{0.257}\\
\cline{2-7}
                                 &300      &3.009   &16.507   &10.336  & 0.572  &0.265 \\
 \cline{2-7}
                                 &200      &3.036   &16.852   &10.401  &0.570   &0.283\\
 \cline{2-7}
                                 &100      &3.062   &16.312   &10.566  &0.567   &0.300\\
\bottomrule
  \end{tabular}
\end{table*}

\begin{figure}[t]
\centerline{
\includegraphics[width=0.8\linewidth]{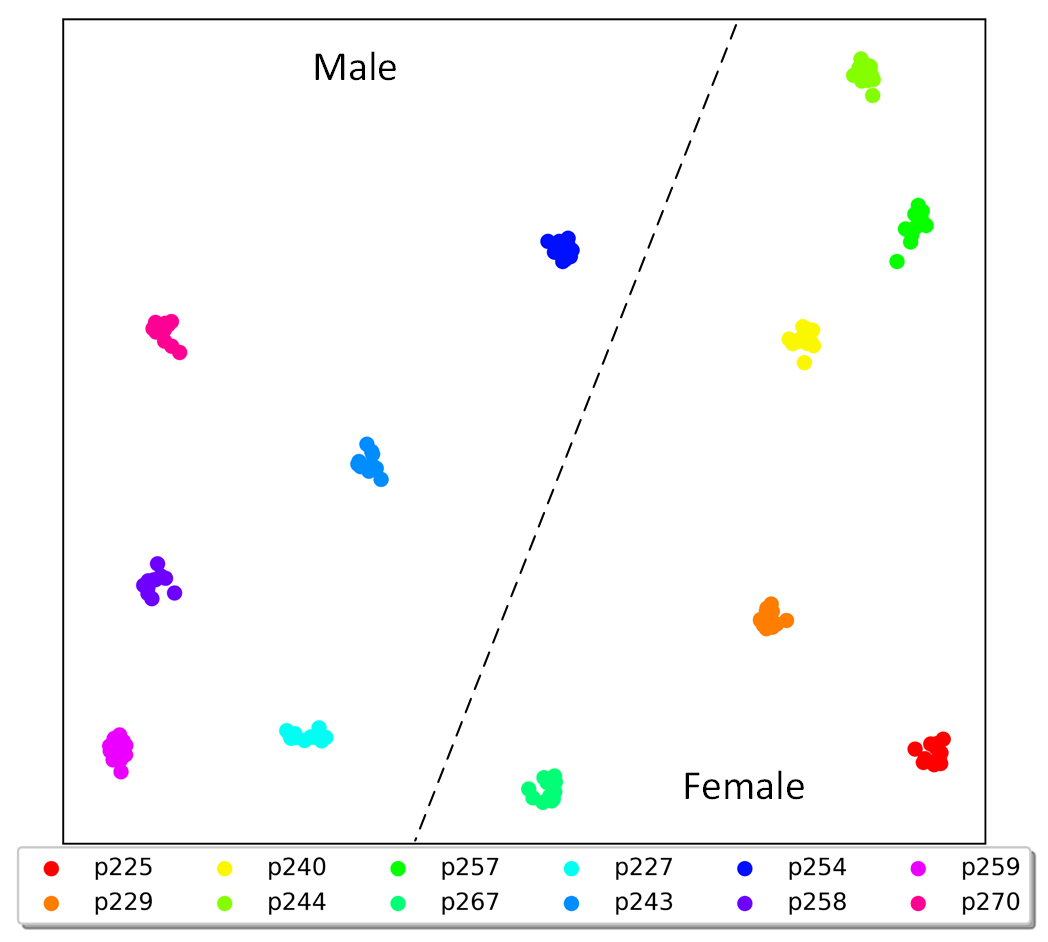}}
  \caption{Visualization of speaker embeddings. Each point represents an utterance and the legend indicates
  different speakers.}
  \label{fig:fig3}
\end{figure}

\begin{figure}[t]
  \centering
  \includegraphics[width=0.75\linewidth]{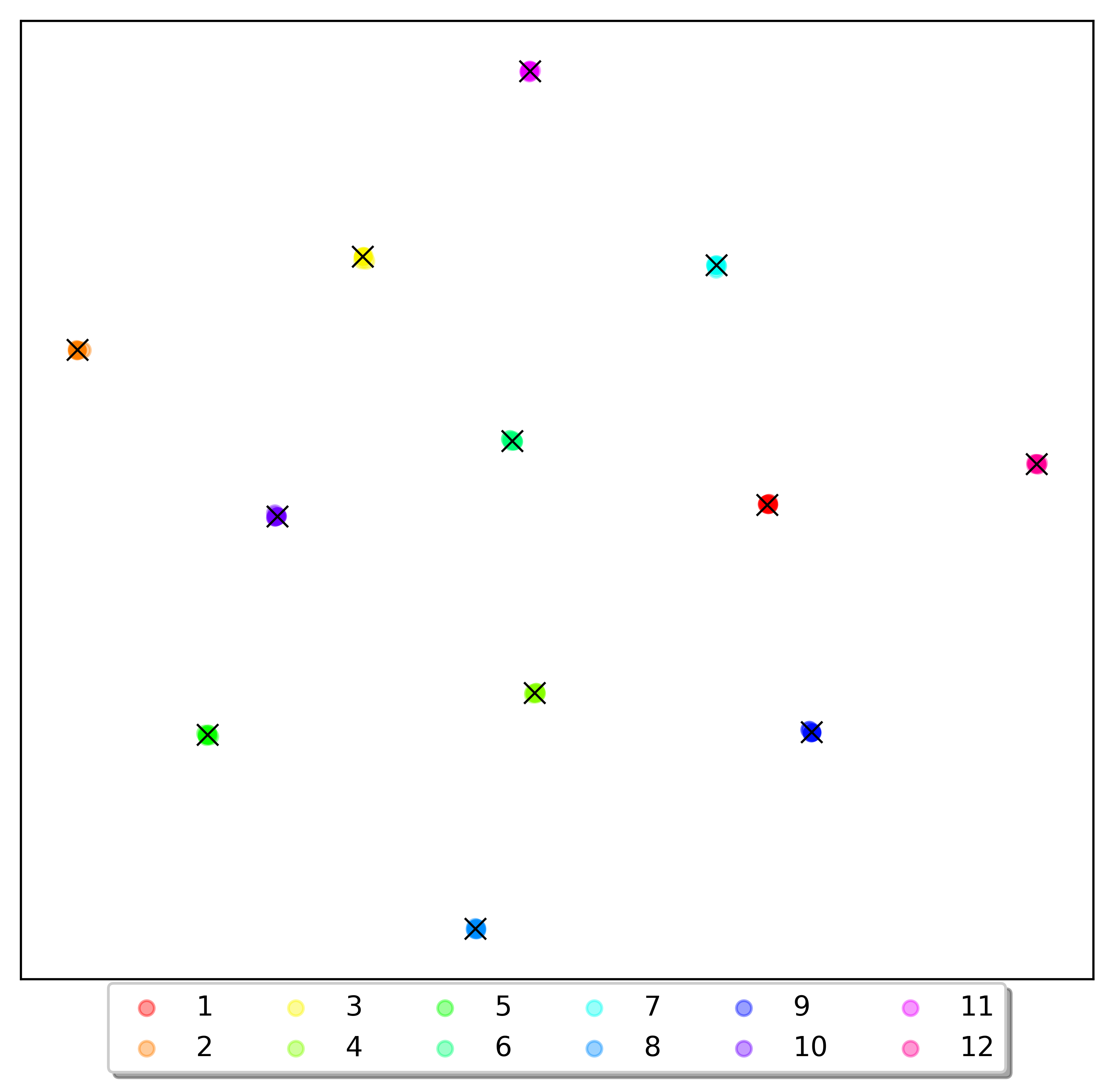}\\
  \centering
  \caption{Visualization of linguistic embeddings. 
  The legend indicates different transcriptions. Each $\times$ symbol  represents the linguistic embedding of a transcription given by the text encoder and
  each point represents the linguistic embedding of an utterance given by the recognition encoder.}
  \label{fig:fig4}
\end{figure}

\subsection{Visualization of hidden representations}
In order to demonstrate that our model can produce disentangled linguistic and speaker representations as we expected,
the extracted linguistic and speaker representations were visualized by t-SNE \cite{maaten2008visualizing}.
12 parallel utterances of 12 speakers in the test set of VCTK were selected and sent into the text encoder, the recognition encoder and the speaker encoder obtained by pre-training.
The linguistic representations $\H^t$ and $\H^r$ given by the text encoder and the recognition encoder were averaged 
along the time axis to get single embedding vector for each utterance.
Then, the  speaker and linguistic embedding vectors of all utterances were projected into a 2-dimensional space by t-SNE and are shown in \figurename~\ref{fig:fig3} and \figurename~\ref{fig:fig4} respectively.

From \figurename~\ref{fig:fig3}, we can see the speaker embeddings from the same speaker were very similar with each other. 
The speaker embeddings of different speakers were also separable according to their genders.
From \figurename~\ref{fig:fig4},  we can see that parallel utterances of different speakers had almost overlapped linguistic representations, which confirmed that
the proposed model can generate speaker-invariant linguistic representations using the recognition encoder.
The linguistic embeddings generated from text inputs were also
located within the clusters of utterances with the same transcriptions, which indicated the
effectiveness of the contrastive loss $\mathcal{L_{CT}}$.

\begin{table*}[t]
 \renewcommand\arraystretch{1.5}
 \caption{Results of ABX preference tests ($\%$) between the proposed method using 100 training utterances and the VCC2018 methods using 500 or 100 training utterances.}\label{tab:pro100}
 \centering
   \begin{tabular}{c l c c c c p{50pt}}
   \toprule
   Conversion Pairs &                    &Proposed (100) & VCC2018 (500) & VCC2018 (100) & \ \ \ \ \ N/P \ \ \ \ \  & $p$-value \\
   \midrule
   \multirow{4}{*}{rms-to-slt}  & Naturalness &37.31         &\textbf{42.31}    &-      & 20.38 & 0.367  \\
                                & Similarity &\textbf{42.69}         &33.85    & -     & 23.46 & 0.103  \\
   \cline{2-7}
                                & Naturalness &\textbf{69.62}         & -      &16.15   &14.23  & $1.36\times10^{-24}$  \\
                                & Similarity &\textbf{75.38}         & -      &13.85   &10.77   &$6.52\times10^{-33}$   \\

   \cline{1-7}
   \multirow{4}{*}{slt-to-rms}  & Naturalness &31.15         &\textbf{39.23}     & - & 29.62 &  0.121 \\
                                & Similarity &\textbf{43.08}         &36.92     & - & 20.00  &  0.268   \\
   \cline{2-7}
                                & Naturalness &\textbf{34.61}         & -        & 27.69   &   37.70    & 0.158 \\
                                & Similarity &\textbf{43.84}         & -        & 26.92    &  29.24     & $1.08\times10^{-3}$   \\

   \bottomrule
   \multicolumn{7}{c}{\emph{N/P} denotes no preference. \emph{100} or \emph{500} indicates the number of non-parallel utterances from each speaker for model training.}
  \end{tabular}
\end{table*}

\begin{table*}[t]
 \renewcommand\arraystretch{1.5}
 \caption{Objective evaluation results of VCC2018 baseline and proposed method on more conversion pairs.}\label{tab:pairs4}
 \centering
  \begin{tabular}{c c c c c c c c}
 \toprule
 Methods            &  &Conversion Pairs & MCD (dB)   & $F_0$ RMSE (Hz) & VUV (\%)  &$F_0$ CORR  &DDUR (s) \\
\midrule
 \multirow{8}{*}{VCC2018} &\multirow{4}{*}{\emph{inter}} &rms-to-slt &3.376  &\textbf{15.042}    &8.222   & \textbf{0.663}  &0.481 \\
  \cline{3-8}
  & &slt-to-rms &3.171  &\textbf{15.771}    &11.382   &\textbf{0.593}  &0.481 \\
  \cline{3-8}
  & &bdl-to-clb &3.669  &15.723    &6.930   &0.667  &0.496 \\
  \cline{3-8}
  & &clb-to-bdl &3.490  &15.199    &11.843   &\textbf{0.657}  &0.496 \\

  \cline{2-8}

  &\multirow{4}{*}{\emph{intra}} &clb-to-slt &3.491  &\textbf{13.997}    &7.250   &\textbf{0.705}  &0.324 \\
  \cline{3-8}
  & &slt-to-clb &3.553  &\textbf{13.013}    &6.250   &\textbf{0.756}  &0.324 \\
  \cline{3-8}
  & & rms-to-bdl &3.312  &\textbf{15.030}    &11.893   &\textbf{0.656}  &0.668 \\
  \cline{3-8}
  & & bdl-to-rms  &3.242  &15.754    &13.458   &0.612  &0.668 \\

  \midrule
   \multirow{8}{*}{Proposed} &\multirow{4}{*}{\emph{inter}} &rms-to-slt &\textbf{3.088}   &16.043   &\textbf{7.898}   &0.624   &\textbf{0.261}\\
  \cline{3-8}
  & &slt-to-rms   &\textbf{2.974}   &16.080   &\textbf{10.327}  &0.581   &\textbf{0.264} \\
  \cline{3-8}
  & &bdl-to-clb  &\textbf{3.150}   &\textbf{15.692}   &\textbf{6.162}  &\textbf{0.672}   &\textbf{0.165} \\
  \cline{3-8}
  & &clb-to-bdl  &\textbf{3.076}   &\textbf{15.078}   &\textbf{11.322}  &0.624   &\textbf{0.191} \\
  \cline{2-8}
  &\multirow{4}{*}{\emph{intra}} &clb-to-slt   &\textbf{3.019}   &15.088   &\textbf{7.128}  &0.662   &\textbf{0.134} \\
  \cline{3-8}
  & &slt-to-clb  &\textbf{3.134}   &14.915   &\textbf{5.600}  &0.698   &\textbf{0.144} \\
  \cline{3-8}
  & & rms-to-bdl   &\textbf{3.157}   &15.192   &\textbf{11.855}  &0.581   &\textbf{0.344} \\
  \cline{3-8}
  & & bdl-to-rms    &\textbf{3.064}   &\textbf{15.214}   &\textbf{10.747}  &\textbf{0.617}   &\textbf{0.359} \\

\bottomrule
  \multicolumn{8}{p{380pt}}{``inter" and ``intra" represent inter-gender and intra-gender conversions respectively. ``slt" and ``clb" are female speakers. ``rms" and ``bdl" are male speakers.}
  \end{tabular}
\end{table*}

\subsection{Evaluation on the amount of training data for fine-tuning}
In this experiment, we gradually reduced the number of training utterances used at the fine-tuning stage
in order to evaluate how the data amount affects the performance of our proposed method.
Five configurations were compared which utilized 500, 400, 300, 200 and 100 training utterances for both source and target speakers respectively.
Their objective performances are summarized in \tablename~\ref{tab:tab4}\footnote{
Since this paper focuses on the acoustic models for voice conversion, the same WaveNet vocoders trained with 500 utterances were used here for all configurations.}.
From \tablename~\ref{tab:tab4}, we can see that
the performance of our proposed method degraded slightly while reducing the number of utterances for fine-tuning. 
Even with only 100 non-parallel utterances, our method still achieved lower MCD than the VCC2018 method in \tablename~\ref{tab:obj} which used 500 training utterances.

Two ABX preference tests were conducted to compare our proposed method using 100 utterances for fine-tuning
with the VCC2018 methods using 500 and 100 utterances respectively.
In each test comparing two methods,
20 test utterances were randomly  selected for each speaker and were converted by both methods to the other speaker.
Then converted utterances were presented to listeners in random order, who were asked to give their preferences in term of both similarity and naturalness. 
At least 13 listeners participated in each test and they were asked to use headphones.
The average preference scores are shown in \tablename~\ref{tab:pro100}.
 From this table, we can see 
that there was no significant difference between our proposed method using 100 utterances and the VCC2018 method using 500 training utterances.
Using the same 100 training utterances, our method achieved significantly better naturalness and similarity than the VCC2018 method,
except the naturalness in slt-to-rms conversion.
These results indicate the advantage of our proposed method when the amount of training data is limited.

\subsection{Evaluation on more conversion pairs}
In order to examine the generalization ability of our proposed method,
experiments were conducted between more conversion pairs.
In additional to the female (slt) and male (rms) speakers used in previous experiments,
another female speaker (clb) and another male speaker (bdl)
 of the CMU ARCTIC dataset were adopted.
The non-parallel datasets were constructed in the same way as the descriptions in Section~\ref{sec:expcond}.
 We compared our proposed method with the VCC2018 baseline.
 The conversion models between two inter-gender speaker pairs and two intra-gender speaker pairs were built
 and evaluated objectively. The results are presented in \tablename~\ref{tab:pairs4}.
We can see that the proposed method obtained consistently better MCD, UVU, DURR metrics than VCC2018 baseline.
In terms of $F_0$ RMSE and $F_0$ CORR, the performance of proposed method was comparable with the baseline.
These results demonstrate the effectiveness of our proposed method on various inter-gender and intra-gender conversion pairs.

\begin{table*}
 \renewcommand\arraystretch{1.5}
 \caption{Objective evaluation results of ablation studies on our proposed method.}\label{tab:tab6}
 \centering
  \begin{tabular}{c l c c c c c c}
 \toprule
Conversion Pairs &  Methods  & MCD (dB)   & $F_0$ RMSE (Hz) & VUV (\%)  &$F_0$ CORR  &DDUR (s)  & PER\\
\midrule
 \multirow{6}{*}{rms-to-slt}  & Proposed  &\textbf{3.088} & \textbf{16.043}   &\textbf{7.898}   &\textbf{0.624}  &\textbf{0.261} &\textbf{10.09}\\
 \cline{2-8}
                                  & $-adv$   &3.256           & 18.426            & 8.985           & 0.499          & 0.406  &10.71\\
 \cline{2-8}
                                 & $-\mathcal{L_{CT}}$ &3.235  &17.065    &8.747  & 0.586  &0.368 & 11.41\\
 \cline{2-8}
                                 & $-text$   &3.613  &22.455    &9.565   & 0.463  &0.488 & 10.45\\
 \cline{2-8}
                                 & $-text-adv$ &4.281  &44.260 & 23.188    & 0.145 & 0.483 & 10.93\\
 \cline{2-8}
                                 & $-$$pre$-$training$ &3.200 & 16.961 & 8.126 & 0.619 &0.593    & 14.81 \\
 \midrule
  \multirow{6}{*}{slt-to-rms}  & Proposed  &\textbf{2.974} & \textbf{16.080}     &\textbf{10.327}   &\textbf{0.581}  &\textbf{0.264} &\textbf{8.84}\\
 \cline{2-8}
                                 & $-adv$  &3.127  &21.227    &11.903  &0.319    &0.374  &9.76\\
 \cline{2-8}
                                 & $-\mathcal{L_{CT}}$  &3.101   &17.170    &10.897  & 0.513  &0.334 &10.36\\
 \cline{2-8}
                                 & $-text$  &3.438  &20.852    &13.197   &0.220  &0.424 &9.25\\
 \cline{2-8}
                                &$-text-adv$ &4.222   & 82.042    & 12.090 & 0.464 & 0.406 &9.21\\
 \cline{2-8}
                                &$-pre$-$training$ &3.120 & 16.866  & 12.359  & 0.551 &0.470   & 17.59 \\
\bottomrule
  \multicolumn{8}{p{390pt}}{``$- adv$'', ``$-\mathcal{L_{CT}}$'' and ``$-text$''
  represent the proposed method without using adversarial training,
  contrastive loss and text inputs respectively. ``$-pre$-$training$'' represents
  the proposed method without using pre-training strategy.}
  \end{tabular}
\end{table*}

\subsection{Ablation studies}
In this section,  ablation studies were conducted to validate the effectiveness of several strategies used in our proposed method,
including the strategies of adversarial training, using text inputs and multi-speaker pre-training.
For investigating the effects of adversarial training, we removed the component of $\C^s$, and the losses of $\mathcal{L_{ADV}}$ and $\mathcal{L_{SC}}$ (as indicated by ``$-adv$" in \tablename~ \ref{tab:tab6}).
For investigating the effects of using text inputs, the contrastive loss $\mathcal{L_{CT}}$ was first removed (i.e., ``$-\mathcal{L_{CT}}$''). Then we further removed the whole text inputs and the text encoder $\E^t$, making the model only learn from acoustic features  (i.e., ``$-text$'').
For investigating the effects of pre-training, the model parameters 
were initialized randomly for fine-tuning  (i.e., ``$-pre$-$training$''). 


\begin{figure*}
  \centering
  \includegraphics[width=0.95\textwidth]{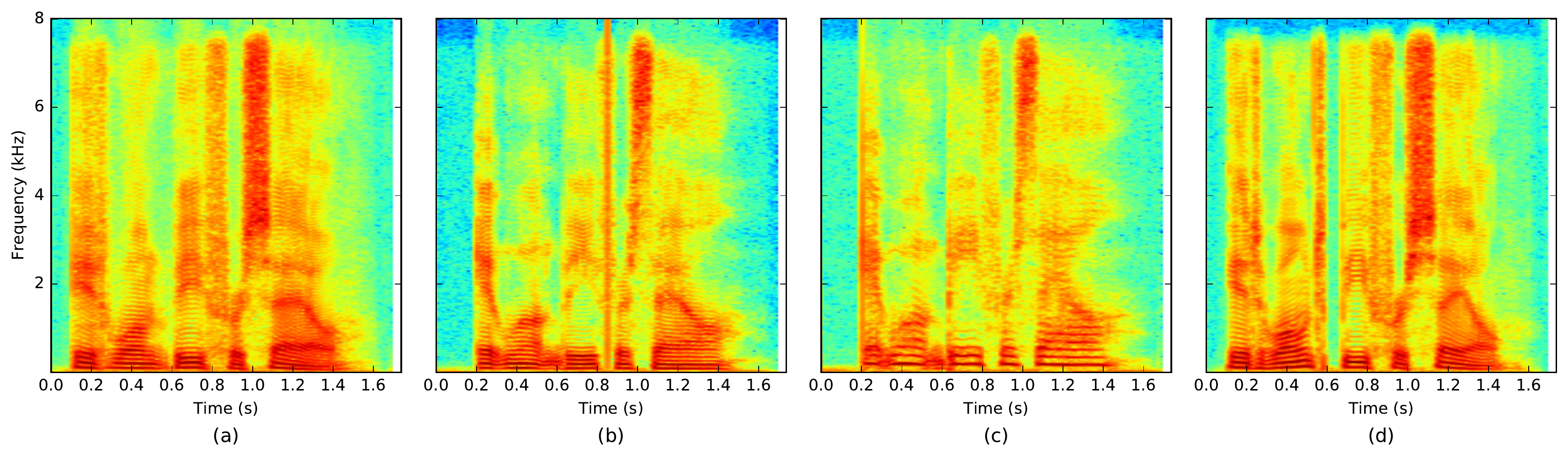}\\
  \centering
  \caption{Spectrograms of (a) the speech  converted by our proposed method,  (b) the  speech converted by ``$-text$'' method, (c) the speech converted by ``$-text-adv$'' method and (d) the natural speech
   of target speaker for a test utterance of slt-to-rms conversion.}\label{fig:fig5}
\end{figure*}
\begin{figure}[t]
  \centering
  \includegraphics[width=0.75\linewidth]{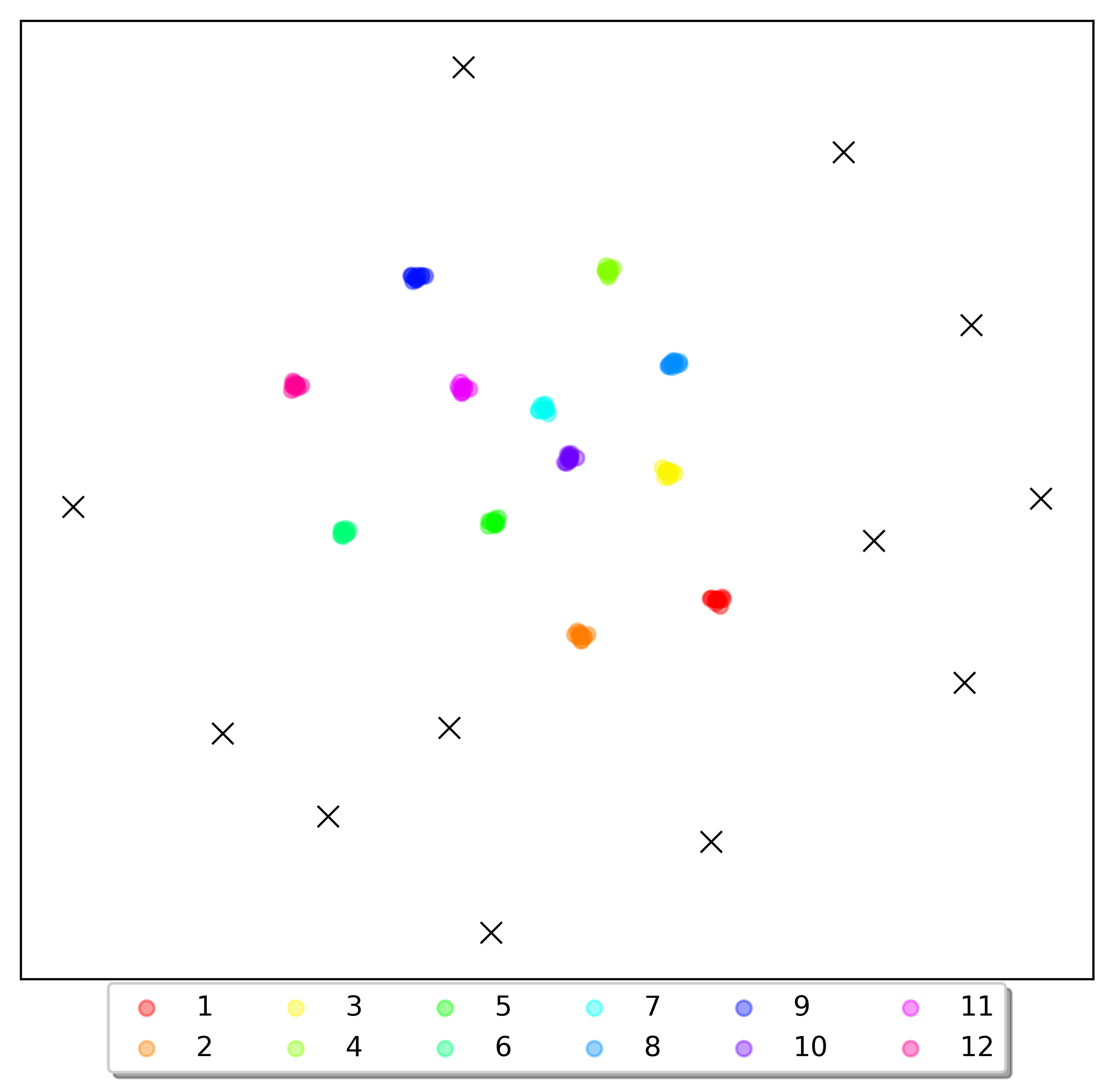}\\
  \caption{Visualization of linguistic embeddings extracted by the model without the contrastive loss $\mathcal{L_{CT}}$.
  The legend indicates different transcriptions. Each $\times$ symbol  represents the linguistic embedding of a transcription given by the text encoder and
  each point represents the linguistic embedding of an utterance given by the recognition encoder.  }
  \label{fig:hidden-contr}
\end{figure}

\begin{figure}[t]
  \centering
  \includegraphics[width=0.75\linewidth]{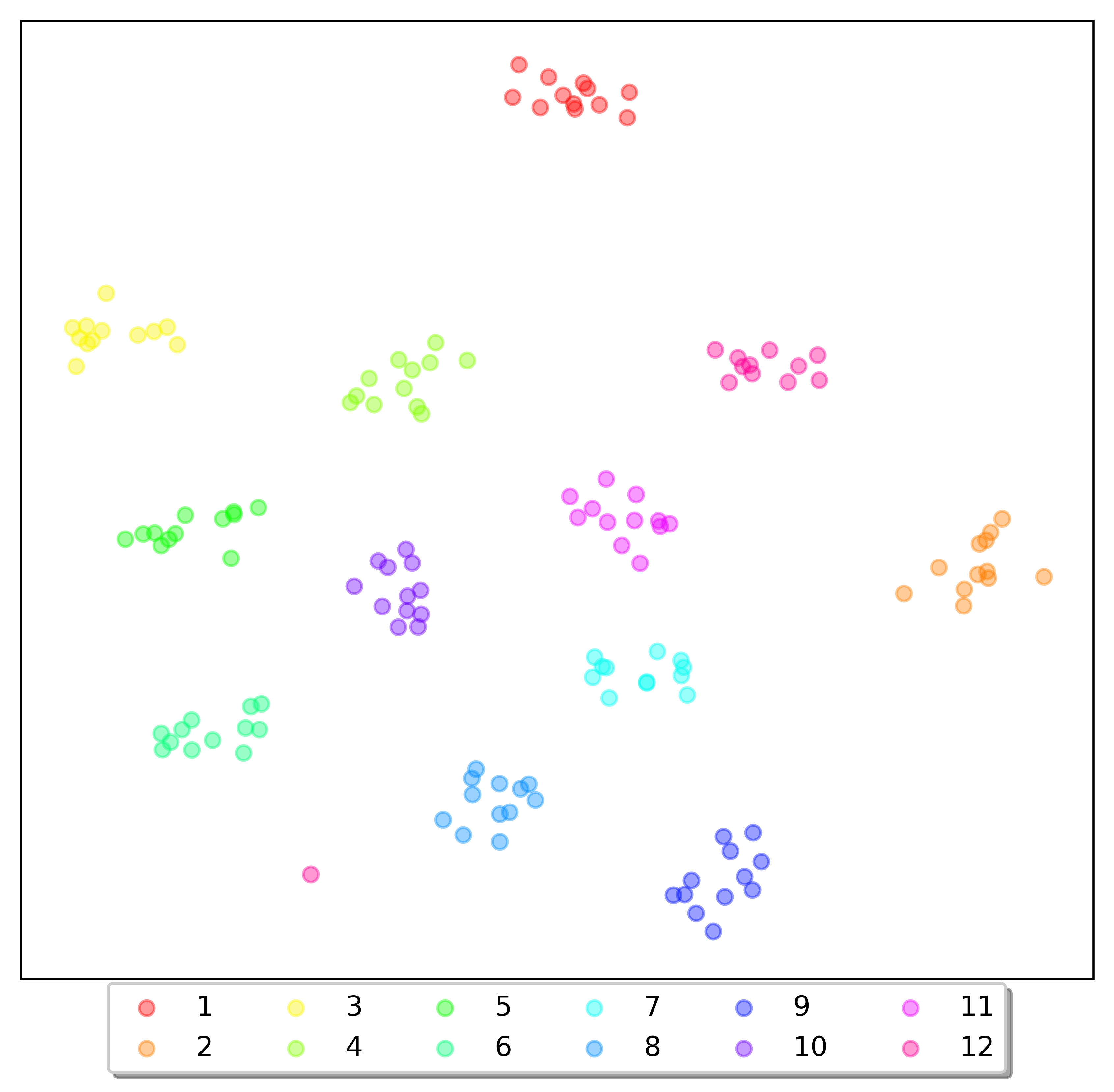}\\
  \caption{Visualization of linguistic embeddings  extracted by the model without both text inputs and adversarial training.
  The legend indicates different transcriptions. 
  Each point represents the linguistic embedding of an utterance given by the recognition encoder.}
  \label{fig:hidden-notextnoadv}
\end{figure}

\tablename~\ref{tab:tab6} shows the objective evaluation results of ablation studies,
which confirmed the effectiveness all proposed strategies.
 In addition to the metrics used in Section~\ref{subsubsec:obj}, the phone error rate (PER) given by the recognition encoder
was employed as shown at the last column of the table.
Without  adversarial training, the performance of proposed method degraded.
After removing contrastive loss $\mathcal{L_{CT}}$, the objective errors increased more seriously than removing adversarial training.
Removing the text inputs and the text encoder caused further degradation.
These results demonstrated that learning linguistic representations jointly with text inputs was crucial in our proposed method. 
The MCD and $F_0$ RMSE metrics increased dramatically if both adversarial training and text inputs were discarded.
In this condition, the model was trained by naive sequence-level auto-encoding on acoustic features.
An informal listening test showed obvious similarity degradation of  converted speech. 
Without the pre-training stage, the PER of the proposed method increased dramatically.
Larger PER means higher risk of mispronunciation in the converted speech.
Our informal listening test indicated  obvious naturalness and intelligibility degradations  of converted speech.
Therefore, it is important to pre-train our model on a large multi-speaker dataset to increase its generalization ability and
to improve the reliability of extracted linguistic representations.



\figurename~\ref{fig:fig5} shows the spectrograms of the speech converted by the proposed, the ``$-text$'' and the  ``$-text-adv$ '' methods together with the spectrogram of
natural target speech for a test utterance of  slt-to-rms conversion.
 As presented in this figure, the proposed method generated the spectrogram which mostly resembled
that of the target.
As shown in \figurename~\ref{fig:fig5} (b), the format patterns of the converted speech without text inputs were inconsistent with those in the target speech. 
If the adversarial training strategy was further discarded, there were serious spectrogram distortions between the converted speech and the target one as shown in \figurename~\ref{fig:fig5} (c),
including a much higher overall pitch of the converted speech than that of the target speech.

\figurename~\ref{fig:hidden-contr} presents the visualization of linguistic embeddings extracted by the proposed model without the contrastive loss $\mathcal{L_{CT}}$.
We can see that the linguistic embeddings extracted from texts scatter around and away from clusters of audio signals, 
 even the same seq2seq decoder was used by text encoder and recognition encoder. 
The linguistic embeddings from the model without both text inputs and adversarial training are also visualized in \figurename~\ref{fig:hidden-notextnoadv}.
From this figure, we can see the
similarities among the utterances of the same transcriptions from different speakers decreased comparing with those in Fig. \ref{fig:fig4}.
This result demonstrated the contributions of text inputs and adversarial training for obtaining disentangled linguistic and speaker representations.

\section{Conclusion}
In this paper, a non-parallel sequence-to-sequence voice conversion method by learning disentangled
linguistic and speaker representations is proposed.
The whole model is built under the framework of encoder-decoder neural networks.
The strategies of using text inputs and adversarial training are adopted for obtaining disentangled linguistic representations.
The model parameters are  pre-trained on a multi-speaker dataset and then fine-tuned on the data of a specific conversion pair.
Experimental results showed that our proposed method surpassed the non-parallel VC method which achieved the
top rank in Voice Conversion Challenge 2018.
The performance of our proposed method was close to the state-of-the-art seq2seq-based parallel VC method.
Ablation studies 
confirmed the effectiveness of adversarial training, using text inputs 
and model pre-training in our proposed method.
Investigating the methods of
one-shot or few-shot voice conversion by improving the prediction of speaker representations in our proposed method
will be our work in future.


%


\ifCLASSOPTIONcaptionsoff
  \newpage
\fi



%
\bibliographystyle{IEEEtran_my}
\bibliography{references}

\begin{thebibliography}{10}
\providecommand{\url}[1]{#1}
\csname url@samestyle\endcsname
\providecommand{\newblock}{\relax}
\providecommand{\bibinfo}[2]{#2}
\providecommand{\BIBentrySTDinterwordspacing}{\spaceskip=0pt\relax}
\providecommand{\BIBentryALTinterwordstretchfactor}{4}
\providecommand{\BIBentryALTinterwordspacing}{\spaceskip=\fontdimen2\font plus
\BIBentryALTinterwordstretchfactor\fontdimen3\font minus
  \fontdimen4\font\relax}
\providecommand{\BIBforeignlanguage}[2]{{%
\expandafter\ifx\csname l@#1\endcsname\relax
\typeout{** WARNING: IEEEtran.bst: No hyphenation pattern has been}%
\typeout{** loaded for the language `#1'. Using the pattern for}%
\typeout{** the default language instead.}%
\else
\language=\csname l@#1\endcsname
\fi
#2}}
\providecommand{\BIBdecl}{\relax}
\BIBdecl

\bibitem{Childers1985Voice}
D.~G. Childers, B.~Yegnanarayana, and K.~Wu, ``Voice conversion: Factors
  responsible for quality,'' in \emph{IEEE International Conference on
  Acoustics, Speech and Signal Processing (ICASSP)}, 1985, pp. 748--751.

\bibitem{Childers1989Voice}
D.~G. Childers, K.~Wu, D.~M. Hicks, and B.~Yegnanarayana, ``Voice conversion,''
  \emph{Speech Communication}, vol.~8, no.~2, pp. 147--158, 1989.

\bibitem{Kain1998Spectral}
A.~Kain, ``Spectral voice conversion for text-to-speech synthesis,'' in
  \emph{IEEE International Conference on Acoustics, Speech and Signal
  Processing (ICASSP)}, vol.~1, 1998, pp. 285--288.

\bibitem{Arslan1999Speaker}
L.~M. Arslan, ``Speaker transformation algorithm using segmental codebooks
  ({STASC}),'' \emph{Speech Communication}, vol.~28, no.~3, pp. 211--226, 1999.

\bibitem{1643640}
C.-H. Wu, C.-C. Hsia, T.-H. Liu, and J.-F. Wang, ``Voice conversion using
  duration-embedded bi-{HMM}s for expressive speech synthesis,'' \emph{IEEE
  Transactions on Audio, Speech, and Language Processing}, vol.~14, no.~4, pp.
  1109--1116, July 2006.

\bibitem{Mohammadi2017An}
S.~H. Mohammadi and A.~Kain, ``An overview of voice conversion systems,''
  \emph{Speech Communication}, vol.~88, pp. 65--82, 2017.

\bibitem{2007Dynamic}
M.~M{\"u}ller, ``Dynamic time warping,'' \emph{Information Retrieval for Music
  and Motion}, pp. 69--84, 2007.

\bibitem{Toda2007Voice}
T.~Toda, A.~W. Black, and K.~Tokuda, ``Voice conversion based on
  maximum-likelihood estimation of spectral parameter trajectory,'' \emph{IEEE
  Transactions on Audio Speech and Language Processing}, vol.~15, no.~8, pp.
  2222--2235, 2007.

\bibitem{Desai2009voice}
S.~Desai, E.~V. Raghavendra, B.~Yegnanarayana, A.~W. Black, and K.~Prahallad,
  ``Voice conversion using artificial neural networks,'' in \emph{IEEE
  International Conference on Acoustics, Speech and Signal Processing
  (ICASSP)}, 2009, pp. 3893--3896.

\bibitem{Desai2010Spectral}
S.~Desai, A.~W. Black, B.~Yegnanarayana, and K.~Prahallad, ``Spectral mapping
  using artificial neural networks for voice conversion,'' \emph{IEEE
  Transactions on Audio Speech and Language Processing}, vol.~18, no.~5, pp.
  954--964, 2010.

\bibitem{Chen2014Voice}
L.-H. Chen, Z.-H. Ling, L.-J. Liu, and L.-R. Dai, ``Voice conversion using deep
  neural networks with layer-wise generative training,'' \emph{IEEE/ACM
  Transactions on Audio Speech and Language Processing}, vol.~22, no.~12, pp.
  1859--1872, 2014.

\bibitem{Sun2015Voice}
L.~Sun, S.~Kang, K.~Li, and H.~Meng, ``Voice conversion using deep
  bidirectional long short-term memory based recurrent neural networks,'' in
  \emph{IEEE International Conference on Acoustics, Speech and Signal
  Processing (ICASSP)}, 2015, pp. 4869--4873.

\bibitem{nakashika2015voice}
T.~Nakashika, T.~Takiguchi, and Y.~Ariki, ``Voice conversion using {RNN}
  pre-trained by recurrent temporal restricted {Boltzmann} machines,''
  \emph{IEEE Transactions on Audio, Speech, and Language Processing}, vol.~23,
  no.~3, pp. 580--587, 2015.

\bibitem{sutskever2014sequence}
I.~Sutskever, O.~Vinyals, and Q.~V. Le, ``Sequence to sequence learning with
  neural networks,'' in \emph{Advances in Neural Information Processing
  Systems}, 2014, pp. 3104--3112.

\bibitem{cho2014learning}
K.~Cho, B.~Van~Merrienboer, C.~Gulcehre, D.~Bahdanau, F.~Bougares, H.~Schwenk,
  and Y.~Bengio, ``Learning phrase representations using {RNN} encoder--decoder
  for statistical machine translation,'' in \emph{Empirical Methods in Natural
  Language Processing}, 2014, pp. 1724--1734.

\bibitem{bahdanau2015neural}
D.~Bahdanau, K.~Cho, and Y.~Bengio, ``Neural machine translation by jointly
  learning to align and translate,'' in \emph{International Conference on
  Learning Representations}, 2015.

\bibitem{luong2015effective}
T.~Luong, H.~Pham, and C.~D. Manning, ``Effective approaches to attention-based
  neural machine translation,'' in \emph{Empirical Methods in Natural Language
  Processing}, 2015, pp. 1412--1421.

\bibitem{8607053}
J.-X. {Zhang}, Z.-H. {Ling}, L.-J. {Liu}, Y.~{Jiang}, and L.-R. {Dai},
  ``Sequence-to-sequence acoustic modeling for voice conversion,''
  \emph{IEEE/ACM Transactions on Audio, Speech, and Language Processing},
  vol.~27, no.~3, pp. 631--644, 2019.

\bibitem{Tanaka2019}
K.~Tanaka, H.~Kameoka, T.~Kaneko, and N.~Hojo, ``{ATTS2S-VC}:
  Sequence-to-sequence voice conversion with attention and context preservation
  mechanisms,'' in \emph{IEEE International Conference on Acoustics, Speech and
  Signal Processing (ICASSP)}, 2019.

\bibitem{zhang2019improving}
J.-X. {Zhang}, Z.-H. {Ling}, Y.~{Jiang}, L.-J. {Liu}, C.~Liang, and L.-R. Dai,
  ``Improving sequence-to-sequence acoustic modeling by adding
  text-supervision,'' in \emph{IEEE International Conference on Acoustics,
  Speech and Signal Processing (ICASSP)}, 2019, pp. 6785--6789.

\bibitem{duxans2006voice}
H.~Duxans, D.~Erro, J.~P{\'e}rez, F.~Diego, A.~Bonafonte, and A.~Moreno,
  ``Voice conversion of non-aligned data using unit selection,'' \emph{TC-STAR
  Workshop on Speech-to-Speech Translation}, 2006.

\bibitem{sundermann2006text}
D.~Sundermann, H.~Hoge, A.~Bonafonte, H.~Ney, A.~Black, and S.~Narayanan,
  ``Text-independent voice conversion based on unit selection,'' in \emph{IEEE
  International Conference on Acoustics, Speech and Signal Processing
  (ICASSP)}, vol.~1, 2006, pp. 81--84.

\bibitem{erro2007frame}
D.~Erro and A.~Moreno, ``Frame alignment method for cross-lingual voice
  conversion,'' in \emph{Annual Conference of the International Speech
  Communication Association (INTERSPEECH)}, 2007, pp. 1969--1972.

\bibitem{erro2010inca}
D.~Erro, A.~Moreno, and A.~Bonafonte, ``{INCA} algorithm for training voice
  conversion systems from nonparallel corpora,'' \emph{IEEE Transactions on
  Audio, Speech, and Language Processing}, vol.~18, no.~5, pp. 944--953, 2010.

\bibitem{kaneko2017parallel}
T.~Kaneko and H.~Kameoka, ``{CycleGAN-VC}: Non-parallel voice conversion using
  cycle-consistent adversarial networks,'' in \emph{European Signal Processing
  Conference (EUSIPCO)}, 2018, pp. 2114--2117.

\bibitem{fang2018high}
F.~Fang, J.~Yamagishi, I.~Echizen, and J.~Lorenzo-Trueba, ``High-quality
  nonparallel voice conversion based on cycle-consistent adversarial network,''
  in \emph{IEEE International Conference on Acoustics, Speech and Signal
  Processing (ICASSP)}, 2018, pp. 5279--5283.

\bibitem{kaneko2019parallel}
T.~Kaneko, H.~Kameoka, K.~Tanaka, and N.~Hojo, ``{CycleGAN-VC2}:improved
  {CycleGAN}-based non-parallel voice conversion,'' in \emph{IEEE International
  Conference on Acoustics Speech and Signal Processing Proceedings}, 2019, pp.
  6820--6824.

\bibitem{nakashika2016non}
T.~Nakashika, T.~Takiguchi, Y.~Minami, T.~Nakashika, T.~Takiguchi, and
  Y.~Minami, ``Non-parallel training in voice conversion using an adaptive
  restricted {Boltzmann} machine,'' \emph{IEEE/ACM Transactions on Audio,
  Speech and Language Processing}, vol.~24, no.~11, pp. 2032--2045, 2016.

\bibitem{sun2016phonetic}
L.~Sun, K.~Li, H.~Wang, S.~Kang, and H.~Meng, ``Phonetic posteriorgrams for
  many-to-one voice conversion without parallel data training,'' in \emph{2016
  IEEE International Conference on Multimedia and Expo (ICME)}, 2016, pp. 1--6.

\bibitem{miyoshi2017voice}
H.~Miyoshi, Y.~Saito, S.~Takamichi, and H.~Saruwatari, ``Voice conversion using
  sequence-to-sequence learning of context posterior probabilities,'' in
  \emph{Annual Conference of the International Speech Communication Association
  (INTERSPEECH)}, 2017.

\bibitem{ljliu2018wav}
L.-J. Liu, Z.-H. Ling, Y.~Jiang, M.~Zhou, and L.-R. Dai, ``{WaveNet} vocoder
  with limited training data for voice conversion,'' in \emph{Annual Conference
  of the International Speech Communication Association (INTERSPEECH)}, 2018,
  pp. 1983--1987.

\bibitem{liu2018voice}
S.~Liu, J.~Zhong, L.~Sun, X.~Wu, X.~Liu, and H.~Meng, ``Voice conversion across
  arbitrary speakers based on a single target-speaker utterance,'' in
  \emph{Annual Conference of the International Speech Communication Association
  (INTERSPEECH)}, 2018, pp. 496--500.

\bibitem{saito2018non}
Y.~Saito, Y.~Ijima, K.~Nishida, and S.~Takamichi, ``Non-parallel voice
  conversion using variational autoencoders conditioned by phonetic
  posteriorgrams and d-vectors,'' in \emph{IEEE International Conference on
  Acoustics, Speech and Signal Processing (ICASSP)}, 2018, pp. 5274--5278.

\bibitem{hsu2016voice}
C.-C. Hsu, H.-T. Hwang, Y.-C. Wu, Y.~Tsao, and H.-M. Wang, ``Voice conversion
  from non-parallel corpora using variational auto-encoder,'' in \emph{2016
  Asia-Pacific Signal and Information Processing Association Annual Summit and
  Conference (APSIPA)}, 2016, pp. 1--6.

\bibitem{hsu2017voice}
C.-C. Hsu, H.-T. Hwang, Y.-C. Wu, Y.~Tsao, and H.-M. Wang, ``Voice conversion
  from unaligned corpora using variational autoencoding {Wasserstein}
  generative adversarial networks,'' in \emph{Annual Conference of the
  International Speech Communication Association (INTERSPEECH)}, 2017, pp.
  3364--3368.

\bibitem{chou2018multi}
J.-c. Chou, C.-c. Yeh, H.-y. Lee, and L.-s. Lee, ``Multi-target voice
  conversion without parallel data by adversarially learning disentangled audio
  representations,'' in \emph{Annual Conference of the International Speech
  Communication Association (INTERSPEECH)}, 2018, pp. 501--505.

\bibitem{denoord2016wavenet}
A.~V. Den~Oord, S.~Dieleman, H.~Zen, K.~Simonyan, O.~Vinyals, A.~Graves,
  N.~Kalchbrenner, A.~W. Senior, and K.~Kavukcuoglu, ``{WaveNet}: A generative
  model for raw audio,'' in \emph{9th ISCA Speech Synthesis Workshop (SSW9)},
  2016, pp. 125--125.

\bibitem{hochreiter1997long}
S.~Hchreiter and J.~Schmidhuber, ``Long short-term memory,'' \emph{Neural
  Computation}, vol.~9, no.~8, pp. 1735--1780, 1997.

\bibitem{polyak2019}
A.~Polyak and L.~Wolf, ``Attention-based {WaveNet} autoencoder for universal
  voice conversion,'' in \emph{IEEE International Conference on Acoustics,
  Speech and Signal Processing (ICASSP)}, 2019.

\bibitem{ocal2019}
O.~Ocal, O.~H. Elibol, G.~Keskin, C.~Stephenson, A.~Thomas, and K.~Ramchandran,
  ``Adversarially trained autoencoders for parallel-data-free voice
  conversion,'' in \emph{IEEE International Conference on Acoustics, Speech and
  Signal Processing (ICASSP)}, 2019.

\bibitem{arik2018neural}
S.~O. Arik, J.~Chen, K.~Peng, W.~Ping, and Y.~Zhou, ``Neural voice cloning with
  a few samples,'' in \emph{Advances in Neural Information Processing Systems},
  2018, pp. 10\,040--10\,050.

\bibitem{jia2018transfer}
Y.~Jia, Y.~Zhang, R.~J. Weiss, Q.~Wang, J.~Shen, F.~Ren, Z.~Chen, P.~Nguyen,
  R.~Pang, I.~L. Moreno \emph{et~al.}, ``Transfer learning from speaker
  verification to multispeaker text-to-speech synthesis,'' in \emph{Advances in
  Neural Information Processing Systems}, 2018, pp. 4485--4495.

\bibitem{nachmani2018fitting}
E.~Nachmani, A.~Polyak, Y.~Taigman, and L.~Wolf, ``Fitting new speakers based
  on a short untranscribed sample,'' in \emph{International Conference on
  Machine Learning}, 2018, pp. 3683--3691.

\bibitem{chan2016listen}
W.~Chan, N.~Jaitly, Q.~Le, and O.~Vinyals, ``Listen, attend and spell: A neural
  network for large vocabulary conversational speech recognition,'' in
  \emph{IEEE International Conference on Acoustics, Speech and Signal
  Processing (ICASSP)}, 2016, pp. 4960--4964.

\bibitem{wang2017tacotron:}
Y.~Wang, R.~J. Skerry-Ryan, D.~Stanton, Y.~Wu, R.~J. Weiss, N.~Jaitly, Z.~Yang,
  Y.~Xiao, Z.~Chen, S.~Bengio \emph{et~al.}, ``Tacotron: Towards end-to-end
  speech synthesis,'' in \emph{Annual Conference of the International Speech
  Communication Association (INTERSPEECH)}, 2017, pp. 4006--4010.

\bibitem{shen2017natural}
J.~Shen, R.~Pang, R.~J. Weiss, M.~Schuster, N.~Jaitly, Z.~Yang, Z.~Chen,
  Y.~Zhang, Y.~Wang, R.~J. Skerry-Ryan \emph{et~al.}, ``Natural {TTS} synthesis
  by conditioning {WaveNet} on mel spectrogram predictions,'' in \emph{IEEE
  International Conference on Acoustics, Speech and Signal Processing
  (ICASSP)}, 2018, pp. 4779--4783.

\bibitem{chopra2005learning}
S.~Chopra, R.~Hadsell, and Y.~LeCun, ``Learning a similarity metric
  discriminatively, with application to face verification,'' in \emph{Computer
  Vision and Pattern Recognition}, 2005, pp. 539--546.

\bibitem{chung2016out}
J.~S. Chung and A.~Zisserman, ``Out of time: automated lip sync in the wild,''
  in \emph{Asian Conference on Computer Vision}, 2016, pp. 251--263.

\bibitem{zhou2018talking}
H.~Zhou, Y.~Liu, Z.~Liu, P.~Luo, and X.~Wang, ``Talking face generation by
  adversarially disentangled audio-visual representation,'' in \emph{AAAI
  Conference on Artificial Intelligence (AAAI)}, 2019.

\bibitem{arctic}
J.~Kominek and A.~W. Black, ``{CMU} {ARCTIC} databases for speech synthesis,''
  \url{http://festvox.org/cmu_arctic/index.html}, 2003, {Lang.} Technol. Inst.,
  Carnegie Mellon Univ., Pittsburgh, PA.

\bibitem{veaux2017cstr}
C.~Veaux, J.~Yamagishi, K.~MacDonald \emph{et~al.}, ``{CSTR} {VCTK} corpus:
  English multi-speaker corpus for cstr voice cloning toolkit,''
  \emph{University of Edinburgh. The Centre for Speech Technology Research
  (CSTR)}, 2017.

\bibitem{chorowski2015attention}
J.~K. Chorowski, D.~Bahdanau, D.~Serdyuk, K.~Cho, and Y.~Bengio,
  ``Attention-based models for speech recognition,'' in \emph{Advances in
  Neural Information Processing Systems}, 2015, pp. 577--585.

\bibitem{radford2015unsupervised}
A.~Radford, L.~Metz, and S.~Chintala, ``Unsupervised representation learning
  with deep convolutional generative adversarial networks,'' in
  \emph{International Conference on Learning Representations}, 2016.

\bibitem{zhang2018forward}
J.-X. Zhang, Z.-H. Ling, and L.-R. Dai, ``Forward attention in
  sequence-to-sequence acoustic modeling for speech synthesis,'' in \emph{IEEE
  International Conference on Acoustics, Speech and Signal Processing
  (ICASSP)}, 2018, pp. 4789--4793.

\bibitem{kingma2014}
D.~Kingma and J.~Ba, ``Adam: A method for stochastic optimization,''
  \emph{Computer Science}, 2014.

\bibitem{Zhizheng2016Merlin}
Z.~Wu, O.~Watts, and S.~King, ``Merlin: An open source neural network speech
  synthesis system,'' in \emph{9th ISCA Speech Synthesis Workshop (SSW9)},
  2016.

\bibitem{Morise2016WORLD}
M.~Morise, F.~Yokomori, and K.~Ozawa, ``{WORLD}: A vocoder-based high-quality
  speech synthesis system for real-time applications,'' \emph{IEICE
  Transactions on Information and Systems}, vol.~99, no.~7, pp. 1877--1884,
  2016.

\bibitem{chappell1998speaker-specific}
D.~T. Chappell and J.~H.~L. Hansen, ``Speaker-specific pitch contour modeling
  and modification,'' in \emph{IEEE International Conference on Acoustics,
  Speech and Signal Processing (ICASSP)}, vol.~2, 1998, pp. 885--888.

\bibitem{Kawahara1999Restructuring}
H.~Kawahara, I.~Masuda-Katsuse, and A.~D. Cheveign{\'e}, ``Restructuring speech
  representations using a pitch-adaptive {time-frequency} smoothing and an
  {instantaneous-frequency} based {F0} extraction: Possible role of a
  repetitive structure in sounds,'' \emph{Speech Communication}, vol.~27, no.
  3–4, pp. 187--207, 1999.

\bibitem{maaten2008visualizing}
L.~v.~d. Maaten and G.~Hinton, ``Visualizing data using t-{SNE},''
  \emph{Journal of Machine Learning Research}, vol.~9, no. Nov, pp. 2579--2605,
  2008.

\end{thebibliography}

%








\end{document}